\documentclass
[superscriptaddress,secnumarabic,amssymb,amsmath,nobibnotes,aps,prd,showkeys,showpacs,nofootinbib,onecolumn]{revtex4}%
\usepackage{graphicx}
\usepackage{epsf}
\usepackage{bm}
\usepackage{amsmath}
\usepackage{amsfonts}
\usepackage{amssymb}%
\providecommand{\U}[1]{\protect\rule{.1in}{.1in}}
\newcommand{\be}{\begin{equation}}
\newcommand{\ee}{\end{equation}}

\newcommand{\mincir}{\raise
-3.truept\hbox{\rlap{\hbox{$\sim$}}\raise4.truept\hbox{$<$}\ }}
\newcommand{\magcir}{\raise
-3.truept\hbox{\rlap{\hbox{$\sim$}}\raise4.truept\hbox{$>$}\ }}

\begin{document}
\title{Dynamics of Chiral Cosmology}
\author{Andronikos Paliathanasis}
\email{anpaliat@phys.uoa.gr}
\affiliation{Institute of Systems Science, Durban University of Technology, Durban 4000,
Republic of South Africa}
\affiliation{Instituto de Ciencias F\'{\i}sicas y Matem\'{a}ticas, Universidad Austral de
Chile, Valdivia 5090000, Chile}

\begin{abstract}
We perform a detailed analysis for the dynamics of Chiral cosmology in a
spatially flat Friedmann-Lema\^{\i}tre-Robertson-Walker universe with a mixed
potential term. The stationary points are categorized in four families.
Previous results in the literature are recovered while new phases in the
cosmological evolution are found. From our analysis we find nine different
cosmological solutions, the eight describe scaling solutions, where the one is
that of a pressureless fluid, while only one de Sitter solution is recovered.

\end{abstract}
\date{\today}
\maketitle

\section{Introduction}

A detailed analysis of the recent cosmological observations
\cite{dataacc1,dataacc2,data1,data2,Hinshaw:2012aka,Ade:2015xua} indicates
that the universe has gone through two acceleration phases during its
evolution. In particular into a late-time acceleration phase which is
attributed to dark energy, and into an early acceleration phase known as
inflation. Inflation was proposed four decades ago \cite{guth} in order to
explain why in large scales the universe appears isotropic and homogeneous.
The inflationary era is described by a scalar field known as the inflaton
which when dominates drives the dynamics of the universe such that the
observations are explained.

In addition, scalar fields have been used to describe the recent acceleration
epoch of the universe, that is, they have been applied as a source of the dark
energy \cite{Ratra}. In scalar field theory the gravitational field equations
remain of second-order with extra degrees of freedom as many as the scalar
fields and corresponding conservation equations \cite{hor1,hor2,hor3}. These
extra degrees of freedom can attribute the geometrodynamical degrees of
freedom provided by invariants to the modification of Einstein-Hilbert action
in the context of modified/alternative theories of gravity
\cite{mod1,mod2,mod3}.

The simplest scalar field theory proposed in the literature is the
quintessence model \cite{Ratra}. Quintessence is described by a minimally
coupled scalar field $\phi\left(  x^{\kappa}\right)  ~$with a potential
function~$V\left(  \phi^{\kappa}\right)  $. The scalar field satisfies the
weak energy condition, i.e. $\rho\geq0,~\rho+p\geq0,$ while the equation of
state parameters $w_{Q}=\frac{p}{\rho}$ is bounded as $\left\vert
w_{Q}\right\vert \leq1$. For some power-law quintessence models, the
gravitational field equations provide finite-time singularities during
inflation leading to chaotic dynamics \cite{sing,page}. On the other hand, for
some kind of potentials the quintessence can describe the late-time
acceleration \cite{udm}.

In the cosmological scenario of a Friedmann-Lema\^{\i}tre-Robertson-Walker
universe (FLRW) exact and analytic solutions of the field equations for
different potentials are presented in
\cite{jdbnew,muslinov,ellis,barrow1,newref2,ref001,ref002} and references
therein. Results of similar analysis on the dynamics of quintessence models
are summarized in the recent review \cite{gen01}. Other scalar field models
which have been proposed in the literature are: phantom fields, Galileon,
scalar tensor, multi-scalar field models and others
\cite{ph1,ph2,ph3,ph4,ph5,ph7,ph8,ph9,ph10,ph11}. Multi-scalar field models
have been used to provide alternative models for the description of inflation
\cite{hy1,hy2,hy3}, such as hybrid inflation, double inflation, $\alpha
$-attractors \cite{hy4,atr1,atr3} and as alternative dark energy models.

Multi-scalar field models which have drawn the attention of cosmologists are,
the quintom model and the Chiral model. A common feature of these two theories
is that they\ are described by two-scalar fields, namely $\phi\left(
x^{\kappa}\right)  $ and $\psi\left(  x^{\kappa}\right)  $. For the quintom
model, one of the two fields is quintessence while the second scalar field is
phantom which means that the energy density of the field can be negative. One
of the main characteristics of quintom cosmology is that the parameter for the
equation of state for the effective cosmological fluid can cross the phantom
divide line more than once \cite{qq1,qq2}. The general dynamics of quintom
cosmology is presented in \cite{qq3}.

In Chiral theory, the two scalar fields have a mixed kinetic term. The two
scalar fields are defined on a two-dimensional space of constant nonvanishing
curvature \cite{atr6,atr7}. That model is inspired by the non-linear sigma
cosmological model \cite{sigm0}. Chiral cosmology is linked with the $\alpha
-$attractor models \cite{atr3}. Exact solutions and for specific cases the
dynamics of Chiral cosmology were studied before in \cite{andimakis}, while
analytic solutions in Chiral cosmology are presented in \cite{2sfand}. In the
latter reference, it was found that pressureless fluid is provided by the
model, consequently, the model can also be seen as an alternative model for
the description of the dark sector of the universe. Last but not least scaling
attractors in Chiral theory were studied in \cite{andimakis,per1}.

In this piece of work we are interested in the evolution of the dynamics for
the gravitational field equations of Chiral cosmology in a spatially flat FLRW
background space. We consider a general scenario where an interaction term for
the two scalar fields exists in the potential term $V\left(  \phi,\psi\right)
$ of the two fields, that is, $V_{,\phi\psi}\neq0$. Specifically, we determine
the stationary points of the cosmological equations and we study the stability
of these points. Each stationary point describes a solution in the
cosmological evolution. Such an analysis is important in order to understand
the general behaviour of the model and to infer about its viability. This
approach has been applied in various gravitational theories with important
results for the viability of specific theories of gravity, see for instance
\cite{dyn1,dyn2,dyn3,dyn4,dyn5,dyn6,dyn7,dyn8,dyn9} and references therein.
From such an analysis we can conclude about for which eras of the cosmological
history can be provided by the specific theory, we refer the reader in the
discussion of \cite{dyn1}. The plan of the paper is as follows.

In Section \ref{sec2} we present the model of our consideration which is that
of Chiral cosmology in a spatially flat FLRW spacetime with a mixed potential
term. We write the field equations which are of second-order. By using the
energy density and pressure variables we observe that the interaction of the
two fields depends on the pressure term. In Section \ref{sec3}, we rewrite the
field equations by using dimensionless variables in the $H-$normalization. We
find an algebraic-differential dynamical system consists of one algebraic
constraint and six first-order ordinary differential equations. We consider a
specific form for the potential in order to reduce dynamical system the system
by one-dimension; and with the use of the constraint equation we end with a
four-dimensional system.

The main results of this work are presented in Section \ref{sec4}. We find the
stationary points of the field equations which form four different families.
The stationary points of family A are those of quintessence, in family B only
the kinetic part of the second scalar field contributes to the cosmological
solutions. On the other hand, the points of family C are those where only the
dynamic part of the second field contributes. Furthermore, for the
cosmological solutions at the points of family D all the components of the
second field contributes to the cosmological fluid. For all the stationary
points we determine the physical properties which describe the corresponding
exact solutions, as also we determine the stability conditions. An application
of this analysis is presented in Section \ref{sec5} with some numerical
results. Moreover, for completeness of our study we present an analytic
solution of the field equations by using previous results of the literature,
from where we can verify the main results of this work. In Section \ref{sec7a}
we discuss the additional stationary points when matter source is included in
the cosmological model. Finally, in Section \ref{sec7} we draw our conclusions.

\section{Chiral cosmology}

\label{sec2}

We consider the gravitational Action Integral to be \cite{2sfand}%
\begin{equation}
S=\int\sqrt{-g}dx^{4}R-\int\sqrt{-g}dx^{4}\left(  \frac{1}{2}g^{\mu\nu}%
H_{AB}\left(  \Phi^{C}\right)  \nabla_{\mu}\Phi^{A}\nabla_{\nu}\Phi
^{B}+V\left(  \Phi^{C}\right)  \right)  \label{ac.01}%
\end{equation}
where $\Phi^{A}=\left(  \phi\left(  x^{\mu}\right)  ,\psi\left(  x^{\mu
}\right)  \right)  $,~$H_{AB}\left(  \Phi^{C}\right)  $ is a second rank
tensor which defines the kinetic energy of the scalar fields, while $V\left(
\Phi^{C}\right)  $ is the potential.

The Action Integral (\ref{ac.01}) describes a interacting two-scalar field
cosmological model where the interaction follows by the potential $V\left(
\Phi^{C}\right)  =V\left(  \phi,\psi\right)  ,$ and the kinetic part.

In this work we assume that $H_{AB}\left(  \Phi^{C}\right)  $ is diagonal and
admits at least one isometry such that (\ref{ac.01})%
\begin{equation}
S=\int\sqrt{-g}dx^{4}R-\int\sqrt{-g}dx^{4}\left(  \frac{1}{2}g^{\mu\nu}\left(
\phi_{;\mu}\phi_{;\nu}+M\left(  \phi\right)  \psi_{;\mu}\psi_{;\mu}\right)
+V\left(  \Phi^{C}\right)  \right)  \label{ac.02}%
\end{equation}
where $M\left(  \phi\right)  _{,\phi}\neq0$ and $M\left(  \phi\right)  \neq
M_{0}\phi^{2}$. In the latter two cases, $H_{AB}\left(  \Phi^{C}\right)  $
describe a two-dimensional flat space and if it is of Lorentzian signature
then it describes the quintom model. Functional of forms of $M\left(
\phi\right)  $ where $H_{AB}\left(  \Phi^{C}\right)  $ is a maximally
symmetric space of constant curvature $R_{0}$, are given by the second-order
differential equation%
\begin{equation}
2M_{,\phi\phi}M-\left(  M_{,\phi}\right)  ^{2}+2M^{2}R_{0}=0. \label{ac.03}%
\end{equation}

A solution of the latter equation is $M\left(  \phi\right)  =M_{0}%
e^{\kappa\phi}$, which can be seen as the general case since new fields can be
defined under coordinate transformations to rewrite the form of $H_{AB}\left(
\Phi^{C}\right)  $.~This is the case of Chiral model that we study in this work.

Variation with respect to the metric tensor of (\ref{ac.01}) provides the
gravitational field equations%
\begin{equation}
G_{\mu\nu}=H_{AB}\left(  \Phi^{C}\right)  \nabla_{\mu}\Phi^{A}\nabla_{\nu}%
\Phi^{B}-g_{\mu\nu}\left(  \frac{1}{2}g^{\mu\nu}H_{AB}\left(  \Phi^{C}\right)
\nabla_{\mu}\Phi^{A}\nabla_{\nu}\Phi^{B}+V\left(  \Phi^{C}\right)  \right)  ,
\label{ac.04}%
\end{equation}
while variation with respect to the fields $\Phi^{A}$ give the Klein-Gordon
vector-equation%
\begin{equation}
g^{\mu\nu}\left(  \nabla_{\mu}\left(  H_{~B}^{A}\left(  \Phi^{C}\right)
\nabla_{\nu}\Phi^{B}\right)  \right)  +H_{~B}^{A}\left(  \Phi^{C}\right)
\frac{\partial V\left(  \Phi^{C}\right)  }{\partial\Phi^{B}}=0. \label{ac.05}%
\end{equation}

According to the cosmological principle, the universe in large scales is
isotropic and homogeneous described by the spatially flat FLRW spacetime with
line element%
\begin{equation}
ds^{2}=-dt^{2}+a^{2}\left(  t\right)  \left(  dx^{2}+dy^{2}+dz^{2}\right)  .
\label{ac.06}%
\end{equation}
where $a\left(  t\right)  $ denotes the scale factor and the Hubble function
is defined as $H\left(  t\right)  =\frac{\dot{a}}{a}$.

For the line element (\ref{ac.06}) and the second-rank tensor $H_{AB}\left(
\Phi^{C}\right)  $ of our consideration the field equations are written as
follows%
\begin{equation}
3H^{2}=\frac{1}{2}\left(  \dot{\phi}^{2}+M\left(  \phi\right)  \dot{\psi}%
^{2}\right)  +V\left(  \phi\right)  +M\left(  \phi\right)  U\left(
\psi\right)  , \label{ac.07}%
\end{equation}%
\begin{equation}
2\dot{H}+3H^{2}=-\left(  \frac{1}{2}\left(  \dot{\phi}^{2}+M\left(
\phi\right)  \dot{\psi}^{2}\right)  -V\left(  \phi\right)  -M\left(
\phi\right)  U\left(  \psi\right)  \right)  , \label{ac.08}%
\end{equation}%
\begin{equation}
\ddot{\phi}+3H\dot{\phi}-\frac{1}{2}M_{,\phi}\dot{\psi}^{2}+V_{,\phi}\left(
\phi\right)  +M_{,\phi}U\left(  \psi\right)  =0, \label{ac.09}%
\end{equation}%
\begin{equation}
\ddot{\psi}+3H\dot{\psi}+\frac{M_{,\phi}}{M}\dot{\phi}\dot{\psi}+U_{,\psi}=0.
\label{ac.10}%
\end{equation}
where we replaced $V\left(  \phi,\psi\right)  =V\left(  \phi\right)  +M\left(
\phi\right)  U\left(  \psi\right)  $ and we have assumed that the fields
$\phi,\psi$ inherit the symmetries of the FLRW space such that $\phi\left(
x^{\mu}\right)  =\phi\left(  t\right)  $ and $\psi\left(  x^{\mu}\right)
=\psi\left(  t\right)  $. At this point we remark that the field equations
(\ref{ac.08})-(\ref{ac.10}) can be produced by the variation principle of the
point-like Lagrangian%
\begin{equation}
\mathcal{L}\left(  a,\dot{a},\phi,\dot{\phi},\psi,\dot{\psi}\right)
=-3a\dot{a}^{2}+\frac{1}{2}a^{3}\left(  \dot{\phi}^{2}+M\left(  \phi\right)
\dot{\psi}^{2}\right)  -a^{3}\left(  V\left(  \phi\right)  +M\left(
\phi\right)  U\left(  \psi\right)  \right)  , \label{ac.10a}%
\end{equation}
while equation (\ref{ac.07}) can be seen as the Hamiltonian constraint of the
time-independent Lagrangian (\ref{ac.10a}).

An equivalent way to write the field equations (\ref{ac.07}), (\ref{ac.08}) is
by defining the quantities%
\begin{equation}
\rho_{\phi}=\frac{1}{2}\dot{\phi}^{2}+V\left(  \phi\right)  ~,~p_{\phi}%
=\frac{1}{2}\dot{\phi}^{2}-V\left(  \phi\right)  , \label{ac.11}%
\end{equation}%
\begin{equation}
\rho_{\psi}=\left(  \frac{1}{2}\dot{\psi}^{2}+U\left(  \psi\right)  \right)
M\left(  \phi\right)  ~,~p_{\psi}=\left(  \frac{1}{2}\dot{\psi}^{2}-U\left(
\psi\right)  \right)  M\left(  \phi\right)  , \label{ac.12}%
\end{equation}
that is,%
\begin{equation}
3H^{2}=\rho_{\phi}+\rho_{\psi}, \label{ac.13}%
\end{equation}%
\begin{equation}
2\dot{H}+3H^{2}=-\left(  p_{\phi}+p_{\psi}\right)  , \label{ac.14}%
\end{equation}%
\begin{equation}
\dot{\rho}_{\phi}+3H\left(  \rho_{\phi}+p_{\phi}\right)  =\dot{\phi}%
\frac{\partial}{\partial\phi}p_{\psi}, \label{ac.15}%
\end{equation}%
\begin{equation}
\dot{\rho}_{\psi}+3H\left(  \rho_{\psi}+p_{\psi}\right)  =-\dot{\phi}%
\frac{\partial}{\partial\phi}p_{\psi}. \label{ac.16}%
\end{equation}

The latter equations give us an interesting observation, since we can write
the interacting functions of the two fields. The interaction models, with
interaction between dark matter and dark energy have been proposed as an
potential mechanism to explain the cosmic coincidence problem and provide a
varying cosmological constant. Some interaction models which have been studied
before in the literature are presented in
\cite{Amendola:2006dg,Pavon:2007gt,Chimento:2009hj,Arevalo:2011hh,an001,an002}
while some cosmological constraints on interacting models can be found in
\cite{in1,in2,in3,in4}.

\section{Dimensionless variables}

\label{sec3}

We consider the dimensionless variables in the $H$-normalization \cite{cop}
\begin{equation}
\dot{\phi}=\sqrt{6}xH~,~V\left(  \phi\right)  =3y^{2}H^{2}~,~\dot{\psi}%
=\frac{\sqrt{6}}{\sqrt{M\left(  \phi\right)  }}zH~,~U\left(  \psi\right)
=\frac{3}{M\left(  \phi\right)  }u^{2}H^{2}\label{ch.01}%
\end{equation}
or%
\begin{equation}
x=\frac{\dot{\phi}}{\sqrt{6}H}~,~y^{2}=\frac{V\left(  \phi\right)  }{3H^{2}%
}~,~z=\frac{\sqrt{M\left(  \phi\right)  }\dot{\psi}}{\sqrt{6}H}~,~u^{2}%
=\frac{M\left(  \phi\right)  U\left(  \psi\right)  }{3H^{2}},
\end{equation}
where the field equations become%
\begin{align}
\frac{dx}{d\tau} &  =\frac{3}{2}x\left(  x^{2}-\left(  1+u^{2}+y^{2}%
-z^{2}\right)  \right)  -\frac{\sqrt{6}}{2}\left(  \lambda y^{2}+\kappa\left(
u^{2}-z^{2}\right)  \right)  ,\label{ch.02}\\
\frac{dy}{d\tau} &  =\frac{3}{2}y\left(  1+x^{2}+z^{2}-y^{2}-u^{2}\right)
+\frac{\sqrt{6}}{2}\lambda xy,\label{ch.03}\\
\frac{dz}{d\tau} &  =\frac{3}{2}z\left(  z^{2}-\left(  1+u^{2}+y^{2}%
-x^{2}\right)  \right)  -\frac{\sqrt{6}}{2}\left(  \kappa xz+\mu u^{2}\right)
,\label{ch.04}\\
\frac{du}{d\tau} &  =\frac{3}{2}u\left(  1+x^{2}+z^{2}-y^{2}-u^{2}\right)
+\frac{\sqrt{6}}{2}u\left(  \kappa x+\mu z\right)  ,\label{ch.05}\\
\frac{d\mu}{d\tau} &  =\sqrt{\frac{3}{2}}\mu\left(  2\mu z\bar{\Gamma}\left(
\mu,\lambda\right)  -\kappa x-2\mu z\right)  ,\label{ch.06}\\
\frac{d\lambda}{d\tau} &  =\sqrt{6}\lambda^{2}x\left(  \Gamma\left(
\lambda\right)  -1\right)  ,\label{ch.07}%
\end{align}
in which
\begin{equation}
\tau=\ln a,~\lambda\left(  \phi\right)  =\frac{V_{,\phi}}{V}~,~\kappa\left(
\lambda\right)  =\frac{M_{,\phi}}{M}~,~\mu\left(  \phi,\psi\right)  =\frac
{1}{\sqrt{M\left(  \phi\right)  }}\frac{U_{,\psi}}{U},~\label{ch.08}%
\end{equation}
and functions $\Gamma\left(  \lambda\right)  ,~\bar{\Gamma}\left(  \mu
,\lambda\right)  $ are defined as%
\begin{equation}
\Gamma\left(  \lambda\right)  =\frac{V_{,\phi\phi}V}{\left(  V_{,\phi}\right)
^{2}}~,~\bar{\Gamma}\left(  \mu,\lambda\right)  =\frac{U_{,\psi\psi}U}{\left(
U_{,\psi}\right)  ^{2}},\label{ch.09}%
\end{equation}
while the constraint equation is%
\begin{equation}
1-x^{2}-y^{2}-z^{2}-u^{2}=0.\label{ch.10}%
\end{equation}

The equation of state parameter for the effective cosmological fluid
$w_{tot},~$is given in terms of the dimensionless parameters as follows%
\begin{equation}
w_{tot}=-1-\frac{2}{3}\frac{\dot{H}}{H^{2}}=x^{2}+z^{2}-y^{2}-u^{2}
\label{ch.11}%
\end{equation}
while we define the variables%
\begin{equation}
\Omega_{\phi}=x^{2}+y^{2}~,~\Omega_{\psi}=z^{2}+u^{2}, \label{ch.12}%
\end{equation}
with equation of state parameters\qquad%
\begin{equation}
w_{\phi}=-1+\frac{2x^{2}}{x^{2}+y^{2}}~,~w_{\psi}=-1+\frac{2z^{2}}{z^{2}%
+u^{2}}. \label{ch.13}%
\end{equation}

At this point it is important to mention that since the two fields interact
that is not the unique definition of the physical variables $\Omega_{\phi}$
and $\Omega_{\psi}$, $w_{\phi}$ and $w_{\psi}$. Moreover, from the constraint
equation (\ref{ch.10}) it follows that the stationary points are on the
surface of a four-dimensional unitary sphere, while the field equations remain
invariant under the transformations $\left\{  y,u\right\}  \rightarrow\left(
-y,-u\right)  $, that is, the variables $\left\{  x,y,z,u\right\}  $ take
values in the following regions $\left\vert x\right\vert \leq1~,~\left\vert
z\right\vert \leq1~,~0\leq y\leq1~\ $and $0\leq u\leq1$.

For the arbitrary functions $V\left(  \phi\right)  ,$ $U\left(  \psi\right)  $
and $M\left(  \phi\right)  $, there are six dependent, namely $\left\{
x,y,z,u,\lambda,\mu\right\}  $, where in general $\kappa=\kappa\left(
\lambda\right)  $, however the dimension of the system can be reduced by one,
if we apply the constraint condition (\ref{ch.10}).

In the following Section, we determine the stationary points for the cases
where $M\left(  \phi\right)  =M_{0}e^{\kappa\phi},~\ V\left(  \phi\right)
=V_{0}e^{\lambda\phi},~$\ and $U\left(  \psi\right)  =U_{0}\psi^{\frac
{1}{\sigma}}$. Consequently, we calculate $\Gamma\left(  \lambda\right)  =1$
and $\bar{\Gamma}\left(  \mu,\lambda\right)  =1-\sigma$ and $\kappa=const$.
Therefore, $\frac{d\lambda}{d\tau}=0$ is satisfied identically and the
dimension of the dynamical system is reduced by one. Therefore we end with the
dynamical system (\ref{ch.02})-(\ref{ch.06}) with constraint (\ref{ch.10}). We
remark that in Chiral model, the kinetic parts of the two fields are defined
on a two-dimensional space of constant curvature.

\section{Dynamical behaviour}

\label{sec4}

The stationary points of the dynamical system have coordinates which make the
rhs of equations (\ref{ch.02})-(\ref{ch.06}) vanish. We categorize the
stationary points into four families. Family A, are the points with
coordinates $\left(  x_{A},y_{A},z_{A},u_{A},\mu_{A}\right)  =\left(
x_{A},y_{A},0,0,0\right)  $ and correspond to the points of the minimally
coupled scalar field cosmology \cite{cop}.

The points with coordinates $\left(  x_{B},y_{B},z_{B},u_{B},\mu_{B}\right)
=\left(  x_{B},y_{B},z_{B},0,\mu_{B}\right)  $ and $z_{B}\neq0$ define the
points of Family B. These points describe physical solutions without any
contribution of the potential $U\left(  \psi\right)  $ to the energy density
of the total fluid source, but only when $\mu_{B}=0$ there is not any
contribution of potential $U\left(  \psi\right)  $ to the dynamics. When
$\mu_{B}=0$, the stationary points are those found before in \cite{andimakis}.

Points of family $C$ have coordinates $\left(  x_{C},y_{C},z_{C},u_{C},\mu
_{C}\right)  =\left(  x_{C},y_{C},0,u_{C},\mu_{C}\right)  ,~u_{C}\neq0$ which
describe exact solutions with no contribution of the kinetic part of the
scalar fields $\psi$. Finally, the points of family D have coordinates of the
form $\left(  x_{C},y_{C},z_{C},u_{C},\mu_{C}\right)  ~$with$~z_{D}u_{D}\neq0$.

Let $P$ be a stationary point of the dynamical system (\ref{ch.02}%
)-(\ref{ch.06}), that is,$~\dot{q}^{A}=f^{A}\left(  q^{B}\right)  $, where
$f^{A}\left(  P\right)  =0$. In order to study the stability properties of the
critical point $P$, we write the linearized system which is $\delta\dot{x}%
^{A}=J_{B}^{A}\delta x^{B}~$where $J_{B}^{A}$ is the Jacobian matrix at the
point $P$, i.e.$~J_{B}^{A}=\frac{\partial f^{A}\left(  P\right)  }{\partial
x^{B}}$. The eigenvalues $\mathbf{e}\left(  P\right)  $ of the Jacobian matrix
determine the stability of the station point. When all the eigenvalues have
negative real part then point $P$ is an attractor and the exact solution at
the point is stable, otherwise the exact solution at the critical point is
unstable and point $P$ is a source, when all the eigenvalues have a positive
real part, or $P$ is a saddle point.

\subsection{Family A}

There are three stationary points which describe cosmological solutions
without any contribution of the second field $\psi$. The points have
coordinates \cite{cop}%
\begin{equation}
A_{1}^{\pm}=\left(  \pm1,0,0,0,0\right)  ~,~A_{2}=\left(  -\frac{\lambda
}{\sqrt{6}},\sqrt{1-\frac{\lambda^{2}}{6}},0,0,0\right)  . \label{ch.14}%
\end{equation}

Points $A_{1}^{\pm}$ describe universes dominated by the kinetic part of the
scalar field $\phi,~$that is by the term $\frac{1}{2}\dot{\phi}^{2}$. The
physical quantities are derived
\[
\left(  w_{tot}\left(  A_{1}^{\pm}\right)  ,w_{\phi}\left(  A_{1}^{\pm
}\right)  ,w_{\psi}\left(  A_{1}^{\pm}\right)  ,\Omega_{\phi}\left(
A_{1}^{\pm}\right)  ,\Omega_{\psi}\left(  A_{1}^{\pm}\right)  \right)
=\left(  1,1,\nexists,1,0\right)  .
\]

Point $A_{2}$ is physically accepted when $\left\vert \lambda\right\vert
<\sqrt{6},$ the physical quantities are calculated%
\[
\left(  w_{tot}\left(  A_{2}\right)  ,w_{\phi}\left(  A_{2}\right)  ,w_{\psi
}\left(  A_{2}\right)  ,\Omega_{\phi}\left(  A_{2}\right)  ,\Omega_{\psi
}\left(  A_{2}\right)  \right)  =\left(  -1+\frac{\lambda^{2}}{3}%
,-1+\frac{\lambda^{2}}{3},\nexists,1,0\right)  .
\]
Therefore, point $A_{2}$ describes a scaling solution. The latter solution is
that of an accelerated universe when $\left\vert \lambda\right\vert <\sqrt{2}$.

In the case of quintessence scalar field cosmology, points $A_{1}^{\pm}$ are
always unstable, while $A_{2}$ is the unique attractor of the dynamical system
when $\left\vert \lambda\right\vert <\sqrt{3}$. However, for the model of our
analysis the stability conditions are different.

In order to conclude for the stability of the stationary points we determine
the eigenvalues of the linearized dynamical system (\ref{ch.02})-(\ref{ch.06})
around to the stationary points. For the points $A_{1}^{\pm}$ it follows%
\begin{align*}
e_{1}\left(  A_{1}^{\pm}\right)   &  =3,\\
e_{2}\left(  A_{1}^{\pm}\right)   &  =\frac{1}{2}\left(  6\pm\sqrt{6}%
\lambda\right)  ,\\
~e_{3}\left(  A_{1}^{\pm}\right)   &  =\frac{1}{2}\left(  6\pm\sqrt{6}%
\kappa\right)  ,\\
~e_{4}\left(  A_{1}^{\pm}\right)   &  =\mp\sqrt{\frac{3}{2}}\kappa,~\\
e_{5}\left(  A_{1}^{\pm}\right)   &  =\mp\sqrt{\frac{3}{2}}\kappa,
\end{align*}
from where we conclude that points $A_{1}^{\pm}~$are saddle points, while the
solutions at points $A_{1}^{\pm}$ are always unstable' because at least one of
the eigenvalues is always positive, i.e. eigenvalue $e_{1}\left(  A_{1}^{\pm
}\right)  >0$.

For the stationary point $A_{2}$ the eigenvalues are derived%
\begin{align*}
e_{1}\left(  A_{2}\right)   &  =\frac{1}{2}\left(  \lambda^{2}-6\right)  ,\\
e_{2}\left(  A_{2}\right)   &  =\lambda^{2}-3,\\
~e_{3}\left(  A_{2}\right)   &  =\frac{1}{2}\kappa\lambda,\\
~e_{4}\left(  A_{2}\right)   &  =\frac{1}{2}\left(  \lambda^{2}-\kappa
\lambda\right)  ,~\\
e_{5}\left(  A_{2}\right)   &  =\frac{1}{2}\left(  \lambda^{2}-6+\kappa
\lambda\right)  ,
\end{align*}
$\ $that is, the exact solution at point $A_{2}$ is always unstable. However.
from the two eigenvalues $e_{1}\left(  A_{2}\right)  ,~e_{2}\left(
A_{2}\right)  $ we can infer that in the surface $\left\{  x,y\right\}  $ of
the phase space the stationary point $A_{2}$ acts like an attractor for
$\left\vert \lambda\right\vert <\sqrt{3}$, which however becomes a saddle
point for the higher-dimensional phase space.

We remark that we determined the stability of the stationary points without
using the constant equation and reducing the dynamical system by
one-dimension. However, by replacing $z^{2}=1-x^{2}-y^{2}-u^{2}$ in the
(\ref{ch.02})-(\ref{ch.06}) we end with a four-dimensional system, from where
we find the same results, that is, the exact solutions at the points
$A_{1}^{\pm}$ and $A_{2}$ are always unstable.

\subsection{Family B}

\begin{figure}[ptb]
\centering\includegraphics[width=0.4\textwidth]{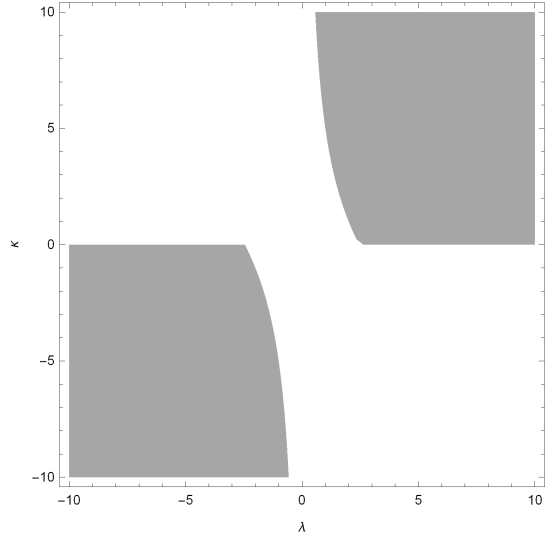} \caption{Region
plot in the space $\left\{  \lambda,\kappa\right\}  $ where points
$\mathbf{B=}\left(  B_{1}^{\pm},B_{2}^{\pm}\right)  $ are real.}%
\label{fig2b1}%
\end{figure}

For $z_{B}\neq0$ and $u_{B}=0,$ we found four stationary points which are%
\begin{align}
B_{1}^{\pm}  &  =\left(  -\frac{\sqrt{6}}{\kappa+\lambda},\sqrt{\frac{\kappa
}{\kappa+\lambda}},\pm\sqrt{\frac{\lambda^{2}+\kappa\lambda-6}{\left(
\kappa+\lambda\right)  ^{2}}},0,0\right)  ,\\
B_{2}^{\pm}  &  =\left(  -\frac{\sqrt{6}}{\kappa+\lambda},\sqrt{\frac{\kappa
}{\kappa+\lambda}},\pm\sqrt{\frac{\lambda^{2}+\kappa\lambda-6}{\left(
\kappa+\lambda\right)  ^{2}}},0,\sqrt{\frac{3}{2}}\frac{\kappa}{\sqrt{\left(
\lambda^{2}+\kappa\lambda-6\right)  }}\right)  ,
\end{align}
which are real and are physically accepted when $\left\{  \kappa
>0,\lambda>\sqrt{6}\right\}  $ or $\left\{  0<\lambda\leq\sqrt{6}%
,~\kappa>\frac{6-\lambda^{2}}{\lambda}\right\}  $ or $\left\{  \lambda
<-\sqrt{6},\kappa<0\right\}  $ or $\left\{  -\sqrt{6}<\lambda<0,\kappa
<\frac{6-\lambda^{2}}{\lambda}\right\}  $. The latter region plots are
presented in Fig. \ref{fig2b1}.

The stationary points have the same physical properties, that is, the points
describe universes with the same physical properties, where the physical
quantities have the following values%
\begin{equation}
w_{tot}\left(  \mathbf{B}\right)  =1-\frac{2\kappa}{\kappa+\lambda}~,~w_{\phi
}\left(  \mathbf{B}\right)  =-1+\frac{12}{6+\kappa\left(  \kappa
+\lambda\right)  }~,~w_{\psi}\left(  \mathbf{B}\right)  =1~,
\end{equation}%
\begin{equation}
\Omega_{\phi}\left(  \mathbf{B}\right)  =1-\Omega_{\psi}\left(  \mathbf{B}%
\right)  ~,~\Omega_{\psi}\left(  \mathbf{B}\right)  =\left\vert \frac
{\lambda\left(  \kappa+\lambda\right)  -6}{\left(  \kappa+\lambda\right)
^{2}}\right\vert .
\end{equation}

\begin{figure}[ptb]
\centering\includegraphics[width=0.9\textwidth]{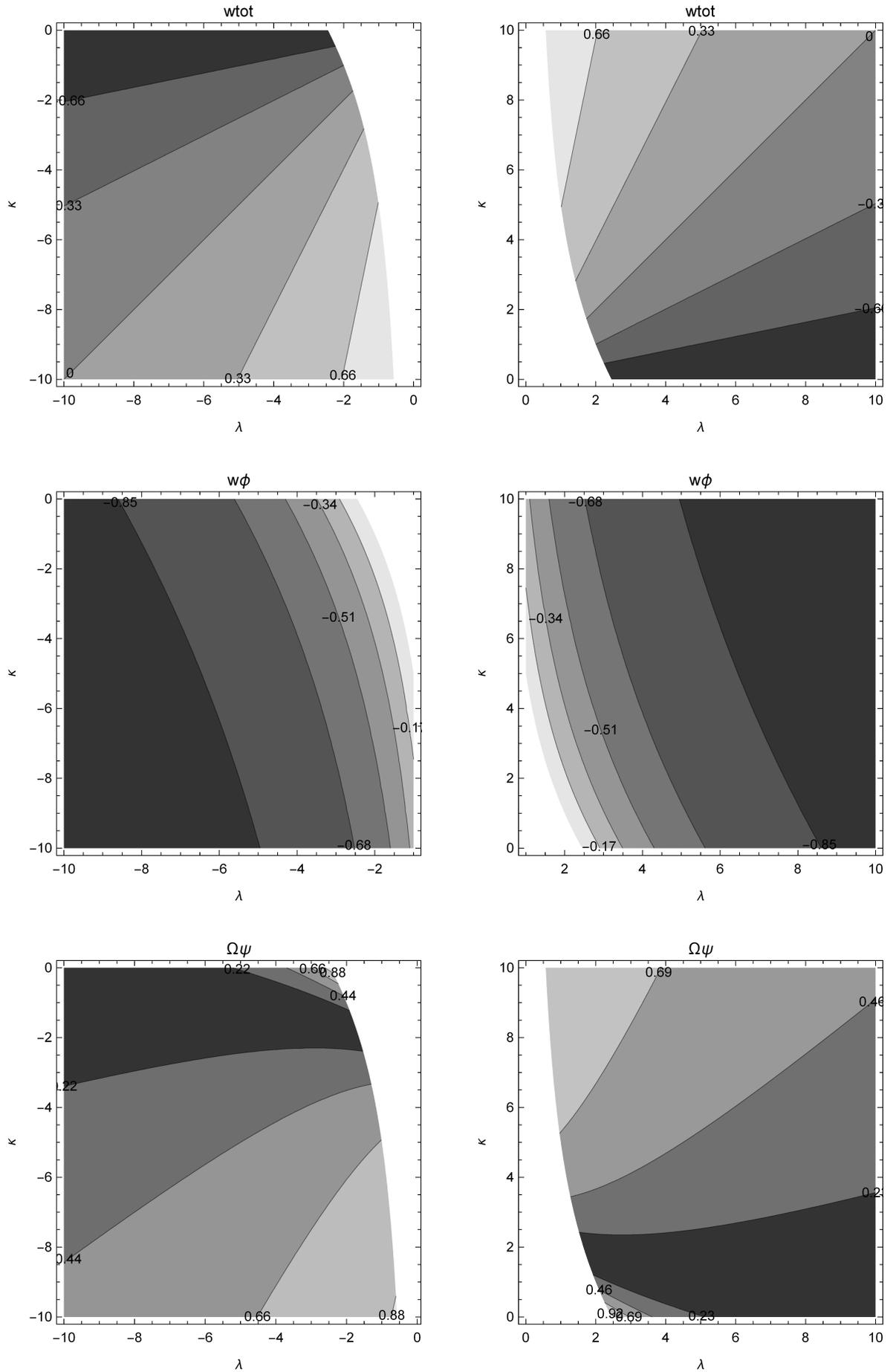}
\caption{Qualitative evolution of the physical variables $w_{tot}\left(
\mathbf{B}\right)  ,~w_{\phi}\left(  \mathbf{B}\right)  \,$\ and $\Omega
_{\psi}\left(  \mathbf{B}\right)  $ of the exact solutions at the critical
points $\mathbf{B=}\left(  B_{1}^{\pm},B_{2}^{\pm}\right)  $ for various
values of the free variables $\left\{  \lambda,\kappa\right\}  $. }%
\label{figb0}%
\end{figure}

From $w_{tot}\left(  \mathbf{B}\right)  $ it follows that the points describe
scaling solutions and the de Sitter universe is recovered only when
$\lambda=0$, which is excluded because for $\lambda=0$, the stationary points
are not real. We continue by studying the stability of the stationary points.
In Fig. \ref{figb0}, we present counter plots for the physical parameters
$w_{tot}\left(  \mathbf{B}\right)  ,~w_{\phi}\left(  \mathbf{B}\right)
\,$\ and $\Omega_{\psi}\left(  \mathbf{B}\right)  $ in the space of variables
$\left\{  \lambda,\kappa\right\}  $.

For the stationary points $B_{1}^{\pm}$ two of the five eigenvalues are
expressed as%
\[
e_{1}\left(  B_{1}^{\pm}\right)  =3\frac{\kappa}{\kappa+\lambda},~e_{2}\left(
B_{1}^{\pm}\right)  =-3\frac{\kappa-\lambda}{\kappa+\lambda},
\]
from where we observe that $e_{1}\left(  B_{1}^{\pm}\right)  >0$ in order for
the points to be real, consequently the exact solutions at the stationary
points $B_{1}^{\pm}$ are unstable.

We use the constraint $z^{2}=1-x^{2}-y^{2}-u^{2}$ such that the dynamical
system is reduced by one-dimension. Thus, for the new four-dimensional system
the eigenvalues of the linearized system around points $B_{1}^{\pm}$ are found%
\begin{align*}
e_{1}\left(  B_{1}^{\pm}\right)   &  =3\frac{\kappa}{\kappa+\lambda}%
,~e_{2}\left(  B_{1}^{\pm}\right)  =-3\frac{\kappa-\lambda}{\kappa+\lambda},\\
e_{3}\left(  B_{1}^{\pm}\right)   &  =-\frac{3\kappa+i\sqrt{3\kappa\left(
4\lambda^{3}+8\kappa\lambda^{2}+4\left(  \kappa^{2}-6\right)  \lambda
-27\kappa\right)  }}{2\left(  \kappa+\lambda\right)  },\\
e_{4}\left(  B_{1}^{\pm}\right)   &  =-\frac{3\kappa-i\sqrt{3\kappa\left(
4\lambda^{3}+8\kappa\lambda^{2}+4\left(  \kappa^{2}-6\right)  \lambda
-27\kappa\right)  }}{2\left(  \kappa+\lambda\right)  },
\end{align*}
from where we conclude again that the exact scaling solutions at points
$B_{1}^{\pm}$ are unstable. In particular points

\begin{figure}[ptb]
\centering\includegraphics[width=0.45\textwidth]{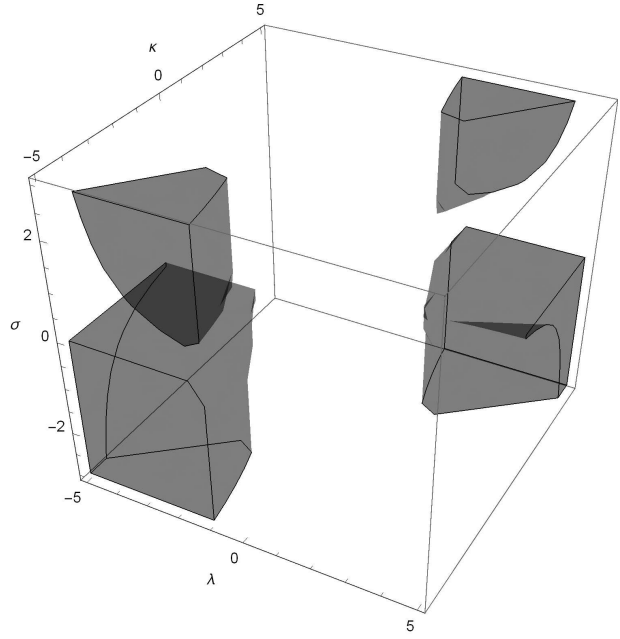}\caption{Region
plot in the space of variabels $\left\{  \kappa,\lambda,\sigma\right\}  $
where the points $B_{2}^{\pm}$ are attractors. }%
\label{figb1}%
\end{figure}

\begin{figure}[ptb]
\centering\includegraphics[width=1\textwidth]{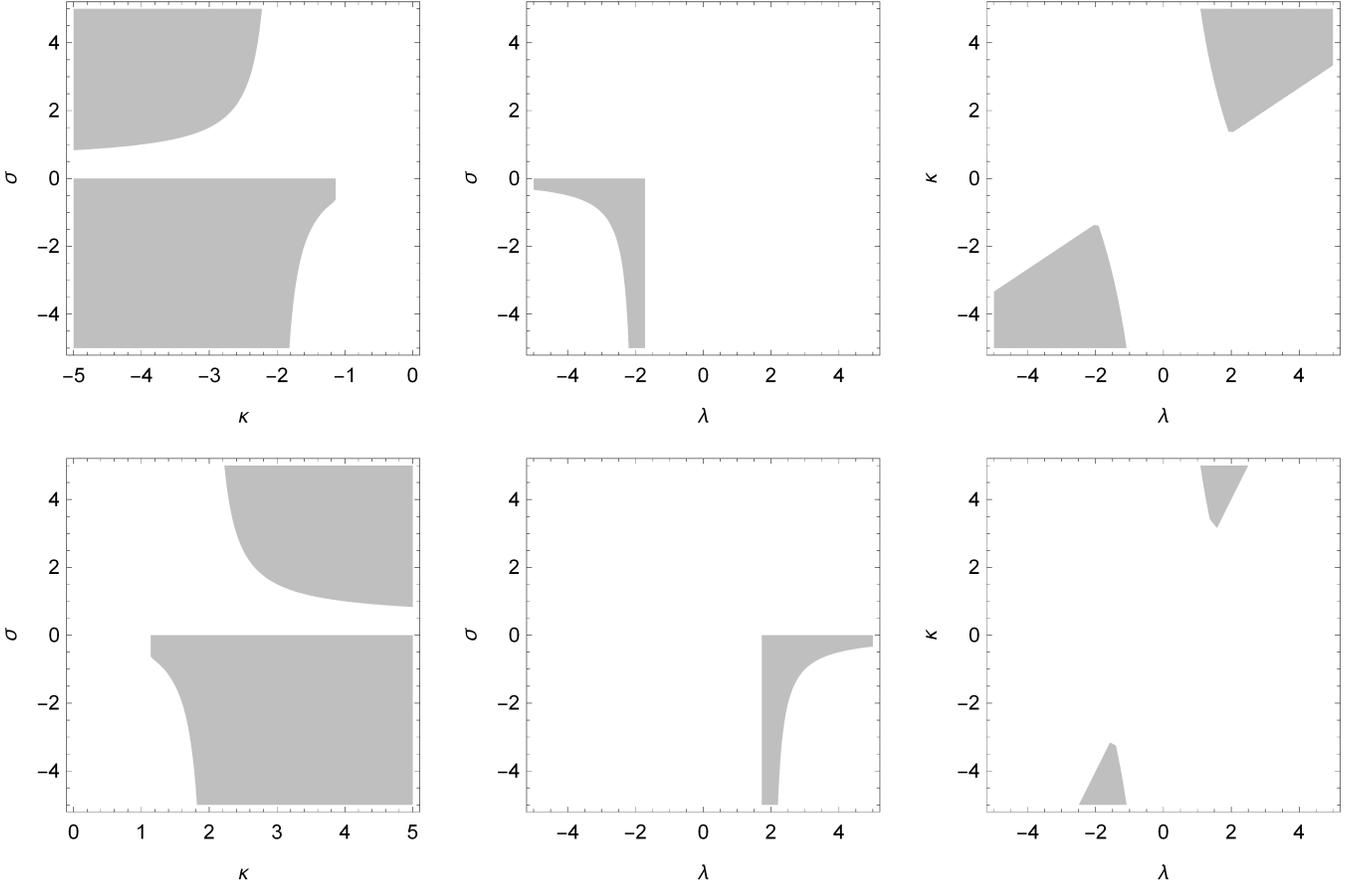} \caption{Region plots
in the the planes $\kappa-\sigma,~\lambda-\sigma$ and $\lambda-\kappa$ where
points $B_{2}^{\pm}$ are attractors. Left figures present the region in the
plane $\kappa-\sigma$ for $\lambda=-2$ and $\lambda=2$; middle figures present
the region in the plane $\lambda-\sigma$, for $\kappa=-2$ and $\kappa-2$ while
right figures are in the plane for $\lambda-\kappa$ for $\sigma=-1$ and
$\sigma=1$.}%
\label{figb2}%
\end{figure}

Similarly, the eigenvalues of the linearized system around the points
$B_{2}^{\pm}$ are calculated \qquad%
\begin{align*}
e_{1}\left(  B_{2}^{\pm}\right)   &  =-3\frac{\kappa}{\kappa+\lambda}%
,~e_{2}\left(  B_{2}^{\pm}\right)  =-3\frac{2\sigma\left(  \kappa
-\lambda\right)  -\kappa}{2\sigma\left(  \kappa+\lambda\right)  },\\
e_{3}\left(  B_{2}^{\pm}\right)   &  =e_{3}\left(  B_{1}^{\pm}\right)
,~e_{4}\left(  B_{2}^{\pm}\right)  =e_{3}\left(  B_{1}^{\pm}\right)  ,
\end{align*}
Hence, we infer that the stationary points $B_{2}^{\pm}$ \ are attractors, and
the exact solutions at the points are stable when the free parameters
$\left\{  \lambda,\kappa,\sigma\right\}  $ are constraints as follows%
\[
\lambda\leq-\sqrt{6}:\left\{  \kappa<\lambda,\sigma<0,\sigma>\frac{\kappa
}{2\left(  \kappa-\lambda\right)  }\right\}  \cup\left\{  \kappa
=\lambda,\sigma<0\right\}  \cup\left\{  \lambda<\kappa<0,\frac{\kappa
}{2\left(  \kappa-\lambda\right)  }<\sigma<0\right\}  ,
\]

\[
-\sqrt{6}<\lambda<-\sqrt{3}:\left\{  \kappa<\lambda,\sigma<0,\sigma
>\frac{\kappa}{2\left(  \kappa-\lambda\right)  }\right\}  \cup\left\{
\kappa=\lambda,\sigma<0\right\}  \cup\left\{  \lambda<\kappa<\frac
{6-\lambda^{2}}{\lambda},\frac{\kappa}{2\left(  \kappa-\lambda\right)
}<\sigma<0\right\}  ,
\]%
\[
\lambda=-\sqrt{3}:\left\{  \kappa<-\sqrt{3},~\sigma<0\right\}  \cup\left\{
\kappa<-\sqrt{3},~\frac{\kappa}{2\left(  \sqrt{3}+\kappa\right)  }%
<\sigma\right\}  ,
\]%
\[
-\sqrt{3}<\lambda<0:\left\{  \kappa<\frac{6-\lambda^{2}}{\lambda}%
,\sigma<0\right\}  \cup\left\{  \kappa<\frac{6-\lambda^{2}}{\lambda}%
,\frac{\kappa}{2\left(  \kappa-\lambda\right)  }<\sigma\right\}  ,
\]%
\[
0<\lambda<\sqrt{3}:\left\{  \kappa>\frac{6-\lambda^{2}}{\lambda}%
,\sigma<0\right\}  \cup\left\{  \kappa>\frac{6-\lambda^{2}}{\lambda}%
,\frac{\kappa}{2\left(  \kappa-\lambda\right)  }<\sigma\right\}  ,
\]%
\[
\lambda=\sqrt{3}:\left\{  \kappa<\sqrt{3},~\sigma<0\right\}  \cup\left\{
\kappa<-\sqrt{3},~-\frac{\kappa}{2\left(  \sqrt{3}-\kappa\right)  }%
<\sigma\right\}  ,
\]%
\[
\sqrt{3}<\lambda<\sqrt{6}:\left\{  \frac{6-\lambda^{2}}{\lambda}%
<\kappa<\lambda,\frac{\kappa}{2\left(  \kappa-\lambda\right)  }<\sigma
<0\right\}  \cup\left\{  \kappa\geq\lambda,\sigma<0\right\}  \cup\left\{
\kappa>\lambda,\frac{\kappa}{2\left(  \kappa-\lambda\right)  }<\sigma\right\}
,
\]%
\[
\lambda\geq\sqrt{6}:\left\{  0<\kappa<\lambda,\frac{\kappa}{2\left(
\kappa-\lambda\right)  }<\sigma<0\right\}  \cup\left\{  \kappa\geq
\lambda,\sigma<0\right\}  \cup\left\{  \kappa>\lambda,\frac{\kappa}{2\left(
\kappa-\lambda\right)  }<\sigma\right\}  .
\]
In Figs. \ref{figb1} and \ref{figb2} we plot the regions where the stationary
points~$B_{2}^{\pm}$ are attractors and the exact solutions on the stationary
points points are stable. 

\subsection{Family C}

The\ stationary points of Family C are two and they have coordinates%
\begin{align}
C_{1}  &  =\left(  -\frac{\kappa}{\sqrt{6}},0,0,\sqrt{1-\frac{\kappa^{2}}{6}%
},0\right)  ,\\
C_{2}  &  =\left(  0,\sqrt{\frac{\kappa}{\kappa-\lambda}},0,\sqrt
{\frac{\lambda}{\lambda-\kappa}},0\right)  .
\end{align}

Point $C_{1}$ is real when $\left\vert \kappa\right\vert \leq\sqrt{6}$ and the
physical quantities of the exact solution at the point are%
\begin{equation}
\left(  w_{tot}\left(  C_{1}\right)  ,w_{\phi}\left(  C_{1}\right)  ,w_{\psi
}\left(  C_{1}\right)  ,\Omega_{\phi}\left(  C_{1}\right)  ,\Omega_{\psi
}\left(  C_{1}\right)  \right)  =\left(  -1+\frac{\kappa^{2}}{3}%
,1,-1,\frac{\kappa^{2}}{6},1-\frac{\kappa^{2}}{6}\right)  .
\end{equation}
Thus, stationary point $C_{1}$ describes a scaling solution. The scaling
solution describes an accelerated universe when $\left\vert \kappa\right\vert
<\sqrt{2}$.

Furthermore, the exact solution at the stationary point $C_{2}$ describes a de
Sitter universe, where the two scalar fields mimic the cosmological constant,
the physical quantities are%
\begin{equation}
\left(  w_{tot}\left(  C_{2}\right)  ,w_{\phi}\left(  C_{2}\right)  ,w_{\psi
}\left(  C_{2}\right)  ,\Omega_{\phi}\left(  C_{2}\right)  ,\Omega_{\psi
}\left(  C_{2}\right)  \right)  =\left(  -1,-1,-1,\frac{\kappa}{\kappa
-\lambda},\frac{\lambda}{\lambda-\kappa}\right)  .
\end{equation}
Point $C_{2}$ is real and physically accepted when $\lambda\kappa<0$, i.e.
$\left\{  \lambda<0,\kappa>0\right\}  $ or $\left\{  \lambda>0,\kappa
<0\right\}  $.

The linearized four-dimensional system around the stationary point $C_{1}$
admits the eigenvalues
\begin{align*}
e_{1}\left(  C_{1}\right)   &  =\frac{\kappa^{2}}{2},\\
e_{2}\left(  C_{1}\right)   &  =-\frac{1}{2}\left(  6-\kappa^{2}\right) \\
e_{3}\left(  C_{1}\right)   &  =2\left(  \kappa^{2}-3\right) \\
e_{4}\left(  C_{1}\right)   &  =\frac{1}{2}\kappa\left(  \kappa-\lambda
\right)
\end{align*}
from where we infer that the exact solution at the stationary point is always
unstable. Specifically, point $C_{1}$ is a saddle point.

\begin{figure}[ptb]
\centering\includegraphics[width=0.45\textwidth]{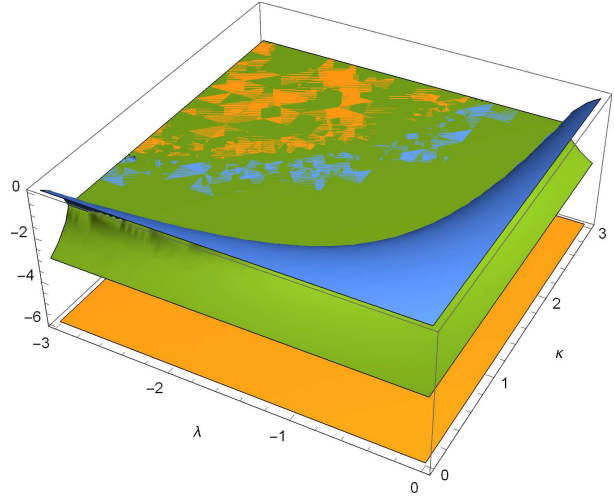}\caption{Qualitative
evolution for the real parts of the nonzero eigenvalues of the linearized
system around the stationary point $C_{2}$. }%
\label{figc}%
\end{figure}

For the stationary point $C_{2}$, we find that one of the eigenvalues of the
linearized system around $C_{2}$ is zero. That eigenvalue corresponds to the
linearize equation (\ref{ch.06}). As far as concerns the other three
eigenvalues we plot numerically their values and we find that they have
negative real parts for all the range of parameters $\left\{  \lambda
,\kappa\right\}  $ where the point exists. In Fig. \ref{figc} we plot the real
parts of the three nonzero eigenvalues of the linearized system.\ Therefore,
we infer that the there exists a four-dimensional stable submanifold around
the stationary point. However, because of the eigenvalues has zero real part
the center manifold theorem (CMT) should be applied.

For simplicity on our calculations we apply the CMT for the five dimensional
system. We find that the variables with nonzero real part on their
eigenvalues, that is, variables $\left\{  x,y,z,u\right\}  $, according to the
CMT theorem are approximated as functions of variable $\mu$ as follows%
\begin{align*}
x  &  =x_{00}\mu^{2}+x_{10}\mu^{3}+x_{20}\mu^{4}+O\left(  \mu^{5}\right)
~,~y=y_{00}\mu^{2}+y_{10}\mu^{3}+y_{20}\mu^{4}+O\left(  \mu^{5}\right)  ,~\\
z  &  =z_{00}\mu^{2}+z_{10}\mu^{3}+z_{20}\mu^{4}+O\left(  \mu^{5}\right)
~,~u=u_{00}\mu^{2}+u_{10}\mu^{3}+u_{20}\mu^{4}+O\left(  \mu^{5}\right)
\end{align*}
where $\left\{  x_{00},y_{00},z_{00},u_{00}\right\}  =\left(  0,0,z_{00}%
,0\right)  $; $x_{10}=-\frac{z_{00}}{\kappa},~y_{10}=\sqrt{\frac{3}{2}}%
\frac{z_{00}}{\sqrt{\kappa^{3}\left(  \kappa-\lambda\right)  }},~$etc.

Hence, the fifth equation, i.e. equation (\ref{ch.02}) is written $\frac{d\mu
}{d\tau}=\alpha\mu^{4}+a_{1}\mu^{5}+O\left(  \mu^{6}\right)  ~$where
$\alpha=\frac{\sqrt{6}\left(  \kappa\lambda-2\left(  \kappa\lambda+3\right)
\sigma\right)  }{2\kappa\lambda+6}z_{00}-\frac{6\kappa\left(  \sqrt
{\lambda\left(  \lambda-\kappa\right)  }\right)  }{2\kappa\lambda+6}u_{10}$.
Therefore, the point is always unstable for $a\neq0$, however~from the
coefficient term $a_{1}\mu^{5}$ we find that the point can be stable.

\subsection{Family D}

The fourth family of stationary points is consists of the following six
stationary points%
\begin{equation}
D_{1}^{\pm}=\left(  -\sqrt{\frac{3}{2}}\frac{1}{\kappa},0,\pm\frac
{\sqrt{\kappa^{2}-3}}{\sqrt{2}\kappa},\frac{1}{\sqrt{2}},0\right)  ,
\end{equation}%
\begin{equation}
D_{2}^{\pm}=\left(  x_{D_{2}},0,\pm z_{D_{2}},\sqrt{1-\left(  x_{D_{2}%
}\right)  ^{2}-\left(  z_{D_{2}}\right)  ^{2}},\mu_{D_{2}}\right)  ,
\end{equation}%
\begin{equation}
D_{3}^{\pm}=\left(  x_{D_{3}},0,\pm z_{D3},\sqrt{1-\left(  x_{D_{3}}\right)
^{2}-\left(  z_{D_{3}}\right)  ^{2}},\mu_{D_{3}}\right)  ,
\end{equation}
with%
\begin{align*}
x_{D_{2}}  &  =-\frac{\kappa^{2}(2\sigma-1)+\sqrt{-4\kappa^{4}\sigma
+\kappa^{4}+4\left(  \kappa^{2}-3\right)  ^{2}\sigma^{2}}+6\sigma}{\sqrt
{6}\kappa(4\sigma-1)},\\
z_{D_{2}}  &  =\frac{\sqrt{-\kappa^{4}(1-2\sigma)^{2}+6\kappa^{2}\sigma\left(
8\sigma^{2}-2\sigma+1\right)  -\sqrt{-4\kappa^{4}\sigma+\kappa^{4}+4\left(
\kappa^{2}-3\right)  ^{2}\sigma^{2}}\left(  \kappa^{2}(2\sigma-1)+24\sigma
^{2}\right)  -144\sigma^{3}}}{2\sqrt{3}\kappa\sqrt{\sigma}(4\sigma-1)},\\
\mu_{D_{2}}  &  =z_{D_{2}}\frac{\sqrt{6}\left(  \kappa^{2}(1-2\sigma
)^{2}+2\sigma\left(  \sqrt{-4\kappa^{4}\sigma+\kappa^{4}+4\left(  \kappa
^{2}-3\right)  ^{2}\sigma^{2}}-6\sigma\right)  \right)  }{\kappa^{2}%
(1-2\sigma)^{2}-24\sigma^{2}},
\end{align*}%
\begin{align*}
x_{D_{3}}  &  =\frac{\kappa^{2}(1-2\sigma)+\sqrt{-4\kappa^{4}\sigma+\kappa
^{4}+4\left(  \kappa^{2}-3\right)  ^{2}\sigma^{2}}-6\sigma}{\sqrt{6}%
\kappa(4\sigma-1)},\\
z_{D_{3}}  &  =\frac{\sqrt{-\kappa^{4}(1-2\sigma)^{2}+6\kappa^{2}\sigma\left(
8\sigma^{2}-2\sigma+1\right)  +\sqrt{-4\kappa^{4}\sigma+\kappa^{4}+4\left(
\kappa^{2}-3\right)  ^{2}\sigma^{2}}\left(  \kappa^{2}(2\sigma-1)+24\sigma
^{2}\right)  -144\sigma^{3}}}{2\sqrt{3}\kappa\sqrt{\sigma}(4\sigma-1)},\\
\mu_{D_{3}}  &  =z_{D_{3}}\frac{\sqrt{6}\left(  2\sigma\left(  \sqrt
{-4\kappa^{4}\sigma+\kappa^{4}+4\left(  \kappa^{2}-3\right)  ^{2}\sigma^{2}%
}+6\sigma\right)  -\kappa^{2}(1-2\sigma)^{2}\right)  }{\kappa^{2}%
(1-2\sigma)^{2}-24\sigma^{2}}\text{. }%
\end{align*}

Points $D_{1}^{\pm}$ describe a scaling solution where the effective fluid is
pressureless, that is, it describes a dust fluid source and the scale factor
is $a\left(  t\right)  =a_{0}t^{\frac{2}{3}}$. The physical parameters of the
exact solution at points $D_{1}^{\pm}$ are%
\begin{equation}
w_{tot}\left(  D_{1}^{\pm}\right)  =0~,~w_{\phi}\left(  D_{1}^{\pm}\right)
=1~,~w_{\psi}\left(  D_{1}^{\pm}\right)  =\frac{3}{3-2\kappa^{2}}~,
\end{equation}%
\begin{equation}
\Omega_{\phi}\left(  D_{1}^{\pm}\right)  =\frac{3}{2\kappa^{2}}~,~\Omega
_{\psi}\left(  D_{1}^{\pm}\right)  =1-\frac{3}{2\kappa^{2}}.
\end{equation}
Remark that points $D_{1}^{\pm}$ are real when $\left\vert \kappa\right\vert
>\sqrt{3}$. The eigenvalues of the four-dimensional linearized system around
the stationary points $D_{1}^{\pm}$ are derived%
\begin{align*}
e_{1}\left(  D_{1}^{\pm}\right)   &  =\frac{3}{2}\\
e_{2}\left(  D_{1}^{\pm}\right)   &  =\frac{3}{2}\left(  \kappa-\lambda\right)
\\
e_{3}\left(  D_{1}^{\pm}\right)   &  =-\frac{3+\sqrt{3\left(  51-16\kappa
^{2}\right)  }}{4}\\
e_{4}\left(  D_{1}^{\pm}\right)   &  =-\frac{3-\sqrt{3\left(  51-16\kappa
^{2}\right)  }}{4}%
\end{align*}
from where we infer that the stationary points $D_{1}^{\pm}$ are always
unstable. Points $D_{1}^{\pm}$ $\ $are saddle points.

\begin{figure}[ptb]
\centering\includegraphics[width=0.8\textwidth]{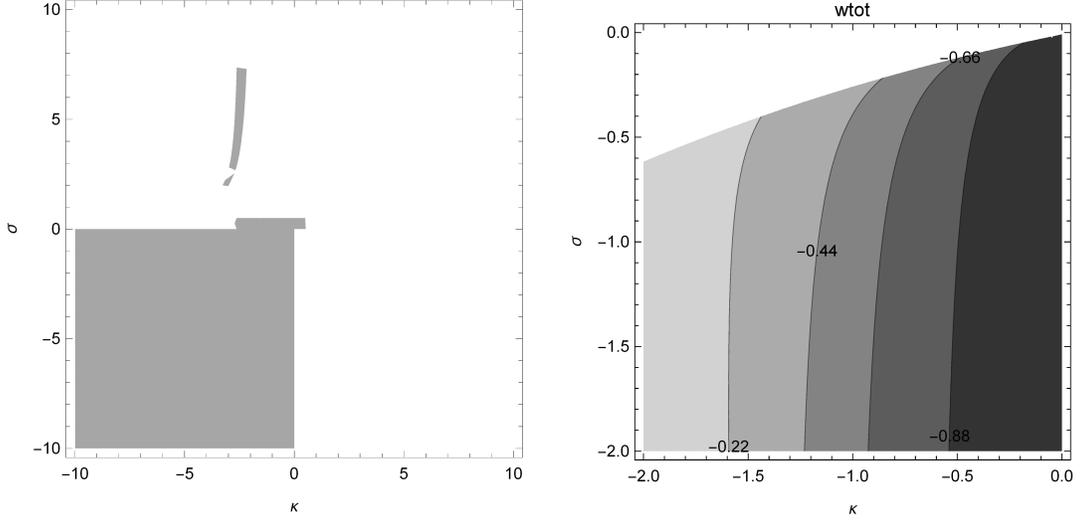} \caption{Left
figure: Region plot in the space $\left\{  \kappa,\sigma\right\}  $ where
points $D_{2}^{\pm}$ are real and physical accepted. Right Figure: Contour
plot of the equation of state parameter for the effective fluid $w_{tot}%
\left(  \kappa,\sigma\right)  $ at the critical points $D_{2}^{\pm}$. }%
\label{figd1}%
\end{figure}

Points $D_{2}^{\pm}$ are real and physically accepted when $\left\{  \sigma
\in\left(  0,\frac{1}{4}\right)  \cup\left(  \frac{1}{4},\frac{1}{2}\right)
,\kappa>\frac{2\sqrt{6}\sigma}{2\sigma-1}\right\}  \cup\left\{  \frac
{2\sqrt{6}\sigma}{1-2\sigma}<\kappa<-\sqrt{6}\sqrt{\frac{2\sigma^{2}%
+\sigma\sqrt{4\sigma-1}}{\left(  1-2\sigma\right)  ^{2}}},\sigma>\frac{1}%
{2}\right\}  $ and $\left\{  \kappa<0,\sigma\,<0\right\}  $ as they are
presented in Fig. \ref{figd1}. The exact solution at the stationary points
describe a scaling solution with values of the equation of state parameter
$w_{tot}\left(  \kappa,\sigma\right)  $ as they presented in Fig. \ref{figd1}.
For the linearized four-dimensional system one of the eigenvalues is
\[
e_{1}\left(  D_{2}^{\pm}\right)  =\frac{A\left(  \kappa,\sigma\right)
(2\kappa\sigma-\kappa-2\lambda\sigma)}{4\kappa\sigma(4\sigma-1)\left(
2\kappa^{2}\sigma-\kappa^{2}+24\sigma^{2}\right)  },
\]
where
\begin{align*}
A\left(  \kappa,\sigma\right)   &  =4\kappa^{4}\sigma^{2}-4\kappa^{4}%
\sigma+\kappa^{4}+48\kappa^{2}\sigma^{3}-12\kappa^{2}\sigma^{2}-6\kappa
^{2}\sigma\\
&  +\sqrt{\left(  2\kappa^{2}\sigma-\kappa^{2}+24\sigma^{2}\right)
^{2}\left(  4\kappa^{4}\sigma^{2}-4\kappa^{4}\sigma+\kappa^{4}-24\kappa
^{2}\sigma^{2}+36\sigma^{2}\right)  }+144\sigma^{3}.
\end{align*}
The other three eigenvalues are only functions of $\kappa,\sigma$, that is
$e_{2,3,4}\left(  D_{2}^{\pm}\right)  =e_{2,3,4}\left(  \kappa,\sigma\right)
$. Numerically, we find that there are not any values of $\left\{
\kappa,\sigma\right\}  $ where the points $D_{2}^{\pm}$ are defined, such that
all the eigenvalues have real part negative, consequently, the stationary
points are always sources and the exact solutions at the stationary points
$D_{2}^{\pm}$ are always unstable.

\begin{figure}[ptb]
\centering\includegraphics[width=0.8\textwidth]{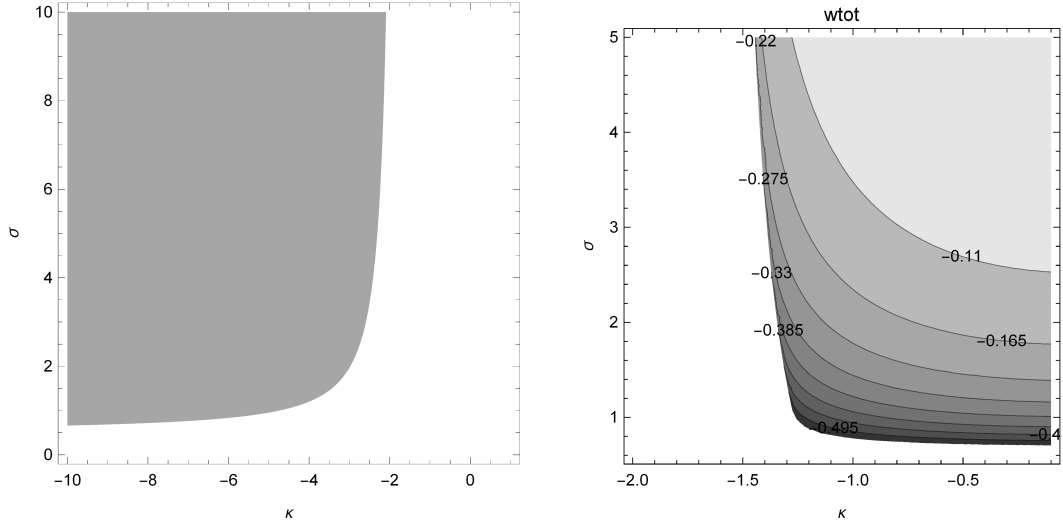} \caption{Left
figure: Region plot in the space $\left\{  \kappa,\sigma\right\}  $ where
points $D_{3}^{\pm}$ are real and physical accepted. Right Figure: Contour
plot of the equation of state parameter for the effective fluid $w_{tot}%
\left(  \kappa,\sigma\right)  $ at the critical points $D_{3}^{\pm}$. }%
\label{figd2}%
\end{figure}

Stationary points $D_{3}^{\pm}$ have similar physical properties with points
$D_{2}^{\pm}$, indeed they describe scaling solutions only. The points are
real and physically accepted in the region $\left\{  \sigma>\frac{1}{2}%
,\kappa<-\sqrt{\frac{6\sigma}{\sqrt{4\sigma-1}-2\sigma}}\right\}  $.

In Fig. \ref{figd2} we present the region in the space $\left\{  \sigma
,\kappa\right\}  $ where the points are defined as also the counter plot of
the equation of state parameter for the effective fluid source which describes
the exact solution at the points $D_{3}^{\pm}$. In a similar way with points
$D_{2}^{\pm}$ we find that there is not any range in the space $\left\{
\kappa,\sigma\right\}  $ where the points are attractors. Consequently, the
stationary points $D_{3}^{\pm}$ $\ $are sources. The main physical results of
the stationary points are summarized in Table \ref{Tab1}.%

%TCIMACRO{\TeXButton{B}{\begin{table}[tbp] \centering}}%
%BeginExpansion
\begin{table}[tbp] \centering
%EndExpansion
\caption{The physical propreties of the stationary models in chiral cosmology}%
\begin{tabular}
[c]{cccccc}\hline\hline
\textbf{Point} & \textbf{Contribution of }$\phi$ & \textbf{Contribution of
}$\psi$ & \textbf{Scaling/de Sitter} & \textbf{Possible }$w_{tot}<-\frac{1}%
{3}$ & \textbf{Stability}\\\hline
$A_{1}$ & Yes only kinetic part & No & Scaling & No & Unstable\\
$A_{2}$ & Yes & No & Scaling & Yes & Unstable\\
$B_{1}^{\pm}$ & Yes & Yes only kinetic part & Scaling & Yes & Unstable\\
$B_{2}^{\pm}$ & Yes & Yes only kinetic part & Scaling & Yes & Can be Stable\\
$C_{1}$ & Yes only kinetic & Yes only potential & Scaling & Yes & Unstable\\
$C_{2}$ & Yes only potential & Yes only potential & de Sitter $\left(
w_{tot}=-1\right)  $ & Always & CMT\\
$D_{1}^{\pm}$ & Yes & Yes & Scaling $\left(  w_{tot}=0\right)  $ & No &
Unstable\\
$D_{2}^{\pm}$ & Yes & Yes & Scaling & Yes & Unstable\\
$D_{3}^{\pm}$ & Yes & Yes & Scaling & Yes & Unstable\\\hline\hline
\end{tabular}
\label{Tab1}%
%TCIMACRO{\TeXButton{E}{\end{table}}}%
%BeginExpansion
\end{table}%
%EndExpansion

\section{Application $\left(  \kappa,\sigma\right)  =\left(  2,\frac{1}%
{2}\right)  $}

\label{sec5}

Consider now the case where $\kappa=2$ and $\sigma=\frac{1}{2}$, while
$\lambda$ is an arbitrary constant. For that consideration, the stationary
points of the dynamical system (\ref{ch.02})-(\ref{ch.06}) have the following
coordinates%
\begin{align*}
\bar{A}_{1}^{\pm}  &  =\left(  \pm1,0,0,0,0\right)  ,~\\
\bar{A}_{2}  &  =\left(  -\frac{\lambda}{\sqrt{6}},\sqrt{1-\frac{\lambda^{2}%
}{6}},0,0,0\right)  ,\\
\bar{B}_{1}^{\pm}  &  =\left(  -\frac{\sqrt{6}}{\lambda+2},\sqrt{\frac
{2}{\lambda+2}},\pm\sqrt{\frac{\left(  \lambda+1\right)  ^{2}-7}{\left(
\lambda+2\right)  ^{2}}},0,0\right)  ,~\\
\bar{B}_{2}^{\pm}  &  =\left(  -\frac{\sqrt{6}}{\lambda+2},\sqrt{\frac
{2}{\lambda+2}},\sqrt{\frac{\left(  \lambda+1\right)  ^{2}-7}{\left(
\lambda+2\right)  ^{2}}},0,2\sqrt{\frac{6}{\left(  \lambda+1\right)  ^{2}-7}%
}\right)  ,\\
\bar{C}_{1}  &  =\left(  -\sqrt{\frac{2}{3}},0,0,\frac{1}{\sqrt{3}},0\right)
,~\\
\bar{C}_{2}  &  =\left(  0,\left(  1-\frac{\lambda}{2}\right)  ^{-1}%
,0,\sqrt{\frac{\lambda}{\lambda-2}}\right)  ,\\
\bar{D}_{1}^{\pm}  &  =\left(  -\frac{1}{2}\sqrt{\frac{3}{2}},0,\frac
{1}{2\sqrt{2}},\frac{1}{\sqrt{2}},0\right)  .
\end{align*}
Points $\bar{A}_{1}^{\pm},~\bar{A}_{2}$ are sources and since they do not
depend on the parameters $\kappa,\sigma$ their physical properties are the
same as before. Recall that point $\bar{A}_{2}$ is real for $\left\vert
\lambda\right\vert <\sqrt{6}$. Stationary points $\mathbf{B}=\left(  \bar
{B}_{1}^{\pm},\bar{B}_{2}^{\pm}\right)  $ exist when $\lambda>\sqrt{7}-1$. The
physical parameters at the points are simplified as follows%
\begin{equation}
w_{tot}\left(  \mathbf{B}\right)  =\frac{\lambda-2}{\lambda+2}~,~w_{\phi
}\left(  \mathbf{B}\right)  =-1+\frac{6}{5\lambda}~,~w_{\psi}\left(
\mathbf{B}\right)  =1~,
\end{equation}%
\begin{equation}
\Omega_{\phi}\left(  \mathbf{B}\right)  =\frac{2\left(  \lambda+5\right)
}{\left(  \lambda+2\right)  ^{2}}~,~\Omega_{\psi}\left(  \mathbf{B}\right)
=1-\frac{2\left(  \lambda+5\right)  }{\left(  \lambda+2\right)  ^{2}}.
\end{equation}
The exact solutions at points $\bar{B}_{1}^{\pm}$ are always unstable.
However, for points $\bar{B}_{2}^{\pm}$ we find that $e_{2}\left(  \bar{B}%
_{2}^{\pm}\right)  >0$ for $\lambda>\sqrt{7}-1$ which means that points
$\bar{B}_{2}^{\pm}$ are sources. The parameter for the equation of state
$w_{tot}\left(  \mathbf{B}\right)  $ is constraint as $\frac{\sqrt{7}-3}%
{\sqrt{7+1}}<w_{tot}\left(  B\right)  <1$, while for $\lambda=2$,
$w_{tot}\left(  \mathbf{B}\right)  =0$ the exact solutions have the scale
factor $a\left(  t\right)  =a_{0}t^{\frac{2}{3}}$, $\ $while for $\lambda=4$,
$w_{tot}\left(  \mathbf{B}\right)  =\frac{1}{3}$, that is $a\left(  t\right)
=a_{0}t^{\frac{1}{2}}$.

Furthermore, stationary point $\bar{C}_{1}$ is a source and describes the
radiation epoch, $w_{tot}\left(  \bar{C}_{1}\right)  =\frac{1}{3}$, on the
other hand, at point $\bar{C}_{2}$ the exact solution is that of de Sitter
universe, the point is real for $\lambda<0\mathbf{.}$Finally, points $\bar
{D}_{1}^{\pm}$ points describe the unstable scaling solutions which describe
the matter dominated era, that is, $w_{tot}\left(  \bar{D}_{1}^{\pm}\right)
=0$.

\begin{figure}[ptb]
\centering\includegraphics[width=0.6\textwidth]{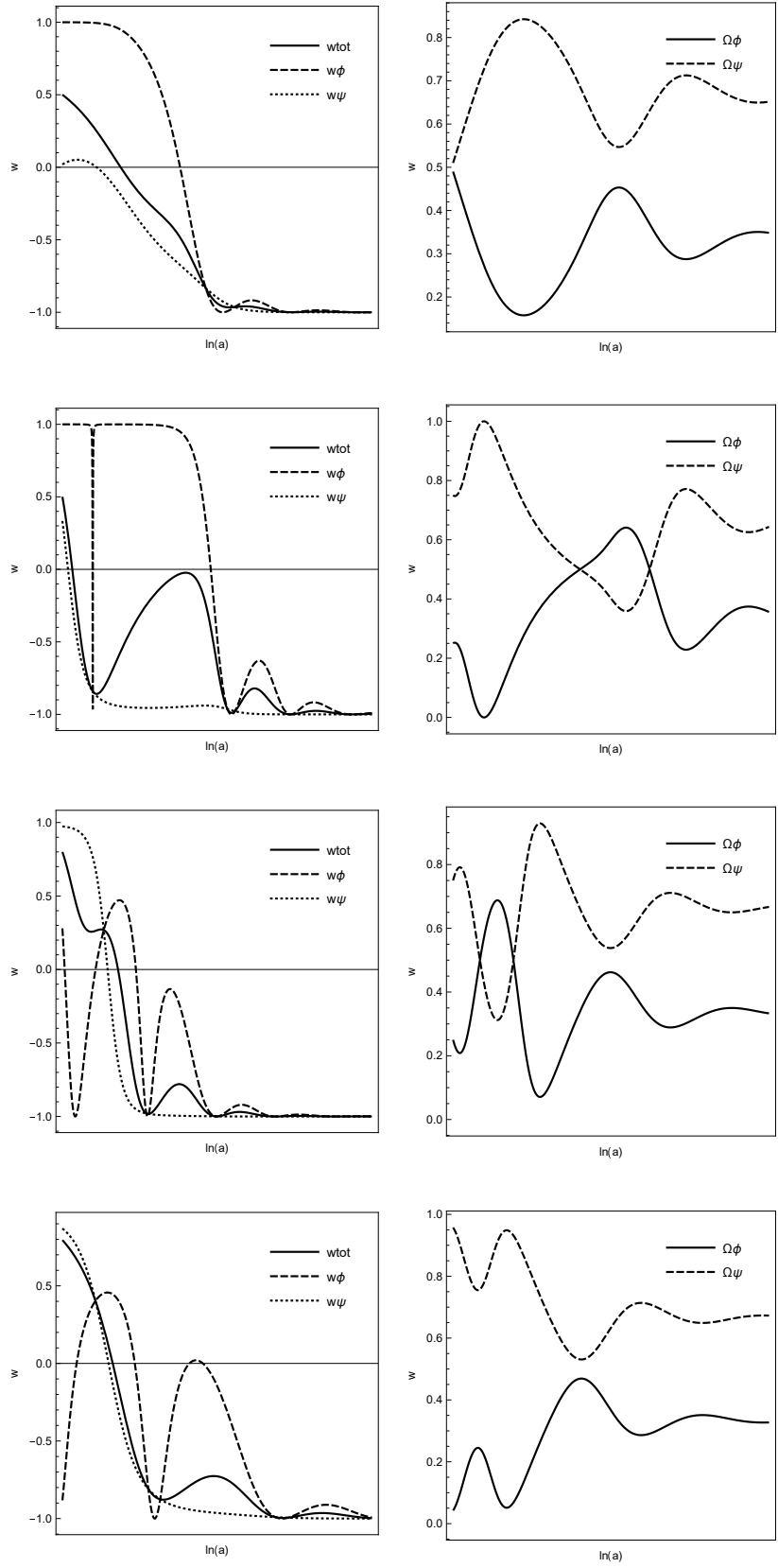}\caption{Evolution
of the physical variables $\left\{  w_{tot},w_{\phi},w_{\psi},\Omega_{\phi
},\Omega_{\psi}\right\}  $ for numerical solutions of the field equations with
$\kappa=2,~\sigma=\frac{1}{2}$ and $\lambda=-4$. The plots are for different
initial conditions $\left(  x\left(  0\right)  ,y\left(  0\right)  ,z\left(
0\right)  ,u\left(  0\right)  ,\mu\left(  0\right)  \right)  $ where
$\mu\left(  0\right)  $ has been chosen to be near to zero, such that the de
Sitter point $\bar{C}_{2}$ to be an attractor. }%
\label{ch2app1}%
\end{figure}

\begin{figure}[ptb]
\centering\includegraphics[width=0.6\textwidth]{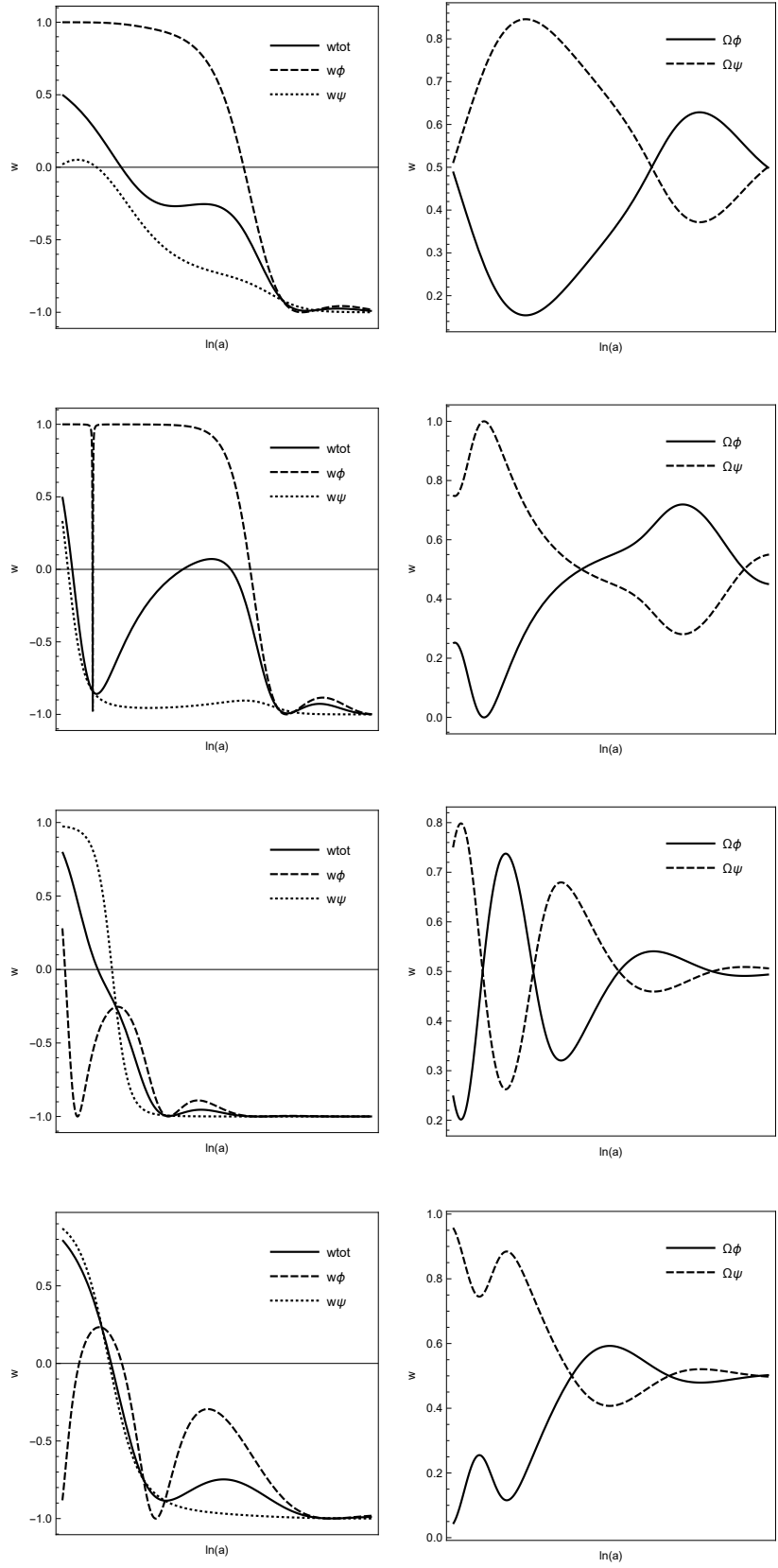}\caption{Evolution
of the physical variables $\left\{  w_{tot},w_{\phi},w_{\psi},\Omega_{\phi
},\Omega_{\psi}\right\}  $ for numerical solutions of the field equations with
$\kappa=2,~\sigma=\frac{1}{2}$ and $\lambda=-2$. The plots are for different
initial conditions $\left(  x\left(  0\right)  ,y\left(  0\right)  ,z\left(
0\right)  ,u\left(  0\right)  ,\mu\left(  0\right)  \right)  $ where
$\mu\left(  0\right)  $ has been chosen to be near to zero, such that the de
Sitter point $\bar{C}_{2}$ to be an attractor. }%
\label{ch2app2}%
\end{figure}

In Figs. \ref{ch2app1} and \ref{ch2app2}, the evolution of the physical
variables $\left\{  w_{tot},w_{\phi},w_{\psi},\Omega_{\phi},\Omega_{\psi
}\right\}  $ is presented for the specific model for $\lambda=-4$ and
$\lambda=-2$ and for different initial conditions for the integration of the
dynamical system (\ref{ch.02})-(\ref{ch.06}). Recall that the de Sitter point
$\bar{C}_{2}$ is a source; however, it admits a four-dimensional stable
manifold when $\mu\rightarrow0$. We observe that in the de Sitter point the
physical parameters $\Omega_{\phi},~\Omega_{\psi}$ are not zero which means
that the all the parts of the potential $V\left(  \phi,\psi\right)  $
contributes to the cosmological fluid. The initial conditions have been
considered such that to describe a wide range of solutions and different
behaviour. The large number of stationary points is observed from the
behaviour of $w_{tot}$, which has various maxima before reach the de Sitter
point. Similarly from the diagram of $\left\{  \Omega_{\phi},~\Omega_{\psi
}\right\}  $, we observe that there is a alternation between the domination of
the two fields.

Consider now the cosmographic parameters $q,~j$ and $s$ which are defined as
\cite{cowei}
\begin{equation}
q\left(  x,y,z,u,\mu;\lambda,\kappa,\sigma\right)  =-1-\frac{\dot{H}}{H^{2}}%
\end{equation}%
\begin{equation}
j\left(  x,y,z,u,\mu;\lambda,\kappa,\sigma\right)  =\frac{\ddot{H}}{H^{3}%
}-3q-2
\end{equation}%
\begin{equation}
s\left(  x,y,z,u,\mu;\lambda,\kappa,\sigma\right)  =\frac{H^{\left(  3\right)
}}{H^{4}}+4j+3q(q+4)+6
\end{equation}

\begin{figure}[ptb]
\centering\includegraphics[width=0.6\textwidth]{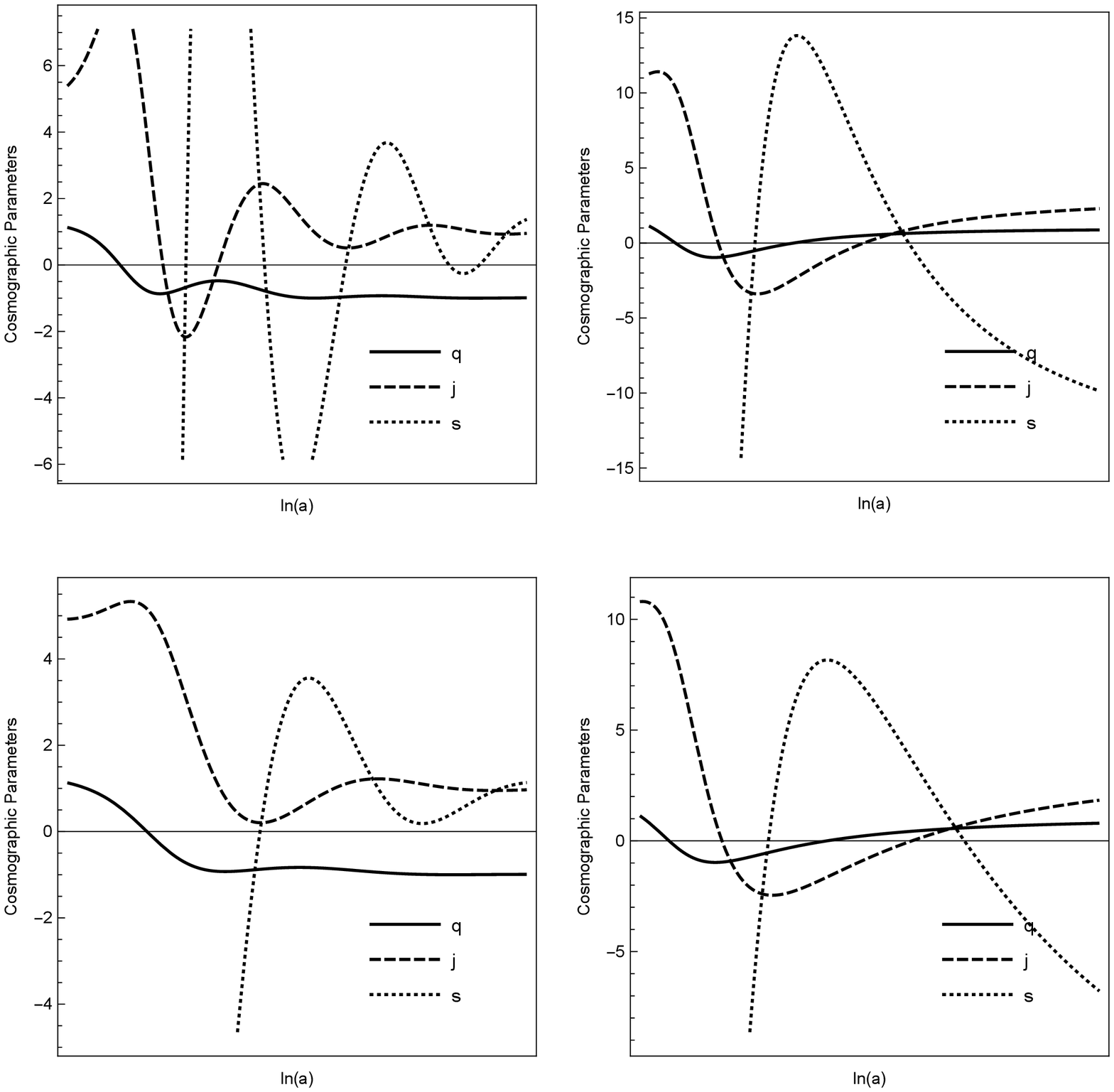}\caption{Qualitative
evolution of the cosmographic parameters $\left\{  q,j,s\right\}  $ for
various values of the free parameters $\left\{  \lambda,\kappa,\sigma\right\}
$. The plots of the first row are for $\left\{  \lambda,\kappa,\sigma\right\}
=\left\{  -4,\pm2,\frac{1}{2}\right\}  $ \ while the plots of the second row
are for $\left\{  \lambda,\kappa,\sigma\right\}  =\left\{  -2,\pm2,\frac{1}%
{2}\right\}  $. \ From the figure we observe that in order the future
attractor to be a de Sitter point then $\kappa>0.$}%
\label{cosmo}%
\end{figure}

In Fig. \ref{cosmo} we present the evolution of the cosmographic parameters
for the application we considered in this example as also for additional
values of the free parameters $\left\{  \lambda,\kappa,\sigma\right\}  $,
while all the plots are for the same initial conditions. Here we present the
qualitative evolution of these parameters, however the cosmographic parameters
as also the free parameters of the theory can be constrained by the
observations \cite{cob1}.

In the following section we continue our analysis by presenting analytic
solutions for the model of our study.

\section{Analytic solution}

\label{sec6}

We consider the point-like Lagrangian
\begin{equation}
\mathcal{L}\left(  a,\dot{a},\phi,\dot{\phi},\psi,\dot{\psi}\right)
=-3a\dot{a}^{2}+\frac{1}{2}a^{3}\left(  \dot{\phi}^{2}+e^{\kappa\phi}\dot
{\psi}^{2}\right)  -a^{3}\left(  V_{0}e^{\lambda\phi}+U_{0}\psi^{\frac
{1}{\sigma}}e^{\kappa\phi}\right)  . \label{ac.100}%
\end{equation}
Analytic solutions of form of Lagrangian (\ref{ac.100}) were presented before
in \cite{2sfand}. By using the results and the analysis of \cite{2sfand} we
present an analytic solutions for specific values of the parameters $\left\{
\lambda,\kappa,\sigma\right\}  $ in order to support the results of the
previous section. Specifically for the free variables we select $\left(
\lambda,\kappa,\sigma\right)  =\left(  -\frac{\sqrt{6}}{2},-\frac{\sqrt{6}}%
{2},\frac{1}{2}\right)  $. These values are not random. In particular, from
the results of \cite{2sfand} it follows that for these specific values the
field equations admit conservation laws and they form a Liouville integrable
dynamical system, such that the field equations can be solved by quadratures.

In order to simplify the field equations and write the analytic solution by
using closed-form functions, we apply the point transformation%
\begin{equation}
a=\left(  xz-\frac{3}{8}y^{2}\right)  ^{\frac{1}{3}}~,~\phi=-2\sqrt{\frac
{2}{3}}\ln\left(  \frac{x}{\sqrt{\left(  xz-\frac{3}{8}y^{2}\right)  }%
}\right)  ~,~\psi=\frac{y}{x} \label{ac.101}%
\end{equation}
such that Lagrangian (\ref{ac.100}) is written as%
\begin{equation}
\mathcal{L}\left(  x,\dot{x},y,\dot{y},z,\dot{z}\right)  =-\frac{4}{3}\dot
{x}\dot{z}-V_{0}x^{2}+\frac{1}{2}\dot{y}^{2}-U_{0}y^{2}. \label{ac.102}%
\end{equation}

In the new coordinates the field equations are%
\begin{equation}
\ddot{x}=0~,~\ddot{y}+2U_{0}y=0~,~\ddot{z}-\frac{3}{2}V_{0}x=0, \label{ac.103}%
\end{equation}
with constraint equation%
\begin{equation}
-\frac{4}{3}\dot{x}\dot{z}+V_{0}x^{2}+\frac{1}{2}\dot{y}^{2}+U_{0}y^{2}=0.
\label{ac.104}%
\end{equation}
Easily, we find the exact solution%
\begin{equation}
x=x_{1}t+x_{0}~,~z=\frac{1}{4}V_{0}x_{1}t^{3}+\frac{3}{4}V_{0}x_{0}t^{2}%
+z_{1}t+z_{0}~, \label{ac.105}%
\end{equation}%
\begin{equation}
y\left(  t\right)  =y_{1}\cos\left(  \sqrt{2U_{0}}t\right)  +y_{2}\sin\left(
\sqrt{2U_{0}}t\right)  \label{ac.106}%
\end{equation}
with constraint condition $V_{0}x_{0}^{2}-\frac{4}{3}x_{1}z_{1}+U_{0}\left(
y_{1}^{2}+y_{2}^{2}\right)  $. \ For $x_{0}=z_{0}=y_{1}=0$, the scale factor
is written $a\left(  t\right)  =\left(  \frac{x_{1}}{4}V_{0}t^{4}+x_{1}%
z_{1}t^{2}-\frac{3}{8}\left(  y_{2}\right)  ^{2}\sin\left(  \sqrt{2U_{0}%
}t\right)  \right)  ^{\frac{1}{3}}.$ It is easy to observe that the present
analytic solution does not provide any de Sitter point. That is in agreement
with the result of the previous section, since for $\lambda=\kappa$, the de
Sitter point $C_{2}$ does not exist. For more general solutions with expansion
eras and de Sitter phases we refer the reader to \cite{2sfand}.

\section{With a matter source}

\label{sec7a}

Let us assume now the presence of an additional pressureless matter source in
field equations with energy density $\rho_{m}$ and let us discuss the
existence of additional stationary points. For a pressureless fluid source the
dimensionless field equations (\ref{ch.02})-(\ref{ch.07}) remain the same,
while the constraint equation (\ref{ch.10}) becomes%
\begin{equation}
\Omega_{m}=1-x^{2}-y^{2}-z^{2}-u^{2}%
\end{equation}
where $\Omega_{m}=\frac{\rho_{m}}{3H^{2}}$, and $0\leq\Omega_{m}\leq1$.

For this model, the stationary points found before exist and give $\Omega
_{m}=0$, while when $\Omega_{m}\neq0$ the additional points exist%
\begin{equation}
E_{1}=\left(  -\sqrt{\frac{3}{2}}\frac{1}{\lambda},\sqrt{\frac{3}{2}}\frac
{1}{\lambda},0,0,0\right)  ~,~E_{2}=\left(  -\sqrt{\frac{3}{2}}\frac{1}%
{\kappa},0,0,\sqrt{\frac{3}{2}}\frac{1}{\kappa},0\right)
\end{equation}%
\begin{equation}
E_{3}=\left(  -\sqrt{\frac{3}{2}}\frac{1}{\kappa},0,z,\sqrt{\frac{3}{2}%
+\kappa^{2}z^{2}}\frac{1}{\kappa},0\right)
\end{equation}
Point $E_{1}$ is physically accepted when $\left\vert \lambda\right\vert
>\sqrt{\frac{3}{2}}$ and describes the tracking solution with $\Omega
_{m}\left(  E_{1}\right)  =1-\frac{3}{\lambda^{2}}$ where the field $\phi$
mimics the ideal gas $\rho_{m},~$that is, $w_{\phi}\left(  E_{1}\right)  =0,$
while the second field $\psi$ does not contribute, i.e. $z\left(
E_{1}\right)  =u\left(  E_{1}\right)  =0$.

For $E_{2}$ we find $\left(  w_{tot}\left(  E_{2}\right)  ,w_{\phi}\left(
E_{2}\right)  ,w_{\psi}\left(  E_{2}\right)  ,\Omega_{\phi}\left(
E_{2}\right)  ,\Omega_{\psi}\left(  E_{2}\right)  \right)  =\left(
0,1,-1,\frac{3}{2\kappa^{2}},\frac{3}{2\kappa^{2}}\right)  $, which means that
it is another tracking tracking solution with $\Omega_{m}=1-\frac{3}%
{\kappa^{2}}$; the point is physically accepted when $\left\vert
\kappa\right\vert \geq\sqrt{\frac{3}{2}}$.

$E_{3}$ does not describe one point, but a family of points on the surface
$u\left(  z\right)  =$ $\sqrt{\frac{3}{2}+\kappa^{2}z^{2}}$ , for $x\left(
E_{3}\right)  =-\sqrt{\frac{3}{2}}\frac{1}{\kappa},$ $y\left(  E_{3}\right)
=\mu\left(  E_{3}\right)  =0$. It describes a tracking solution, that is
$w_{tot}\left(  E_{3}\right)  =0$, with physical parameters%
\begin{equation}
\left(  w_{tot}\left(  E_{3}\right)  ,w_{\phi}\left(  E_{3}\right)  ,w_{\psi
}\left(  E_{3}\right)  ,\Omega_{\phi}\left(  E_{3}\right)  ,\Omega_{\psi
}\left(  E_{3}\right)  \right)  =\left(  0,1,-\frac{3}{4+3\kappa^{2}z^{2}%
},\frac{3}{2\kappa^{2}},2z^{2}+\frac{3}{2\kappa^{2}}\right)  ,
\end{equation}
while the point is physically accepted when \ $\left\vert \kappa\right\vert
\geq\sqrt{\frac{3}{2}}$ and $\left\vert z\right\vert \leq\frac{1}{2}%
\sqrt{2-\frac{3}{\kappa^{2}}}$. When $z\left(  E_{3}\right)  =0$, then $E_{3}$
reduces to $E_{2}$. What it is important, to mention is that the stability
analysis for all the previous points changes, since we made use of the
constraint equation (\ref{ch.10}).

\section{Conclusions}

\label{sec7}

In this work we performed a detailed study of the dynamics for a two scalar
field model with a mixed potential term known as Chiral model. The purpose of
our analysis was to study the cosmological evolution of that specific model as
also the cosmological viability of the model and which epochs of the
cosmological evolution can be described by the Chiral model.

For the scalar field potential we assumed that it is of the form $V\left(
\phi,\psi\right)  =V_{0}e^{\lambda\phi}+U_{0}\psi^{\frac{1}{\sigma}}%
e^{\kappa\phi}$. For this consideration and without assuming the existence of
additional matter source, we found four families of stationary points which
provide nine different cosmological solutions. Eight of the cosmological
solutions are scaling solutions which describe spacetimes with a a perfect
fluid with a constant equation of state parameter $w\left(  P\right)  $. One
of the scaling solutions describes a universe with a stiff matter,~$w\left(
P\right)  =1$, another scaling solution correspond to a universe with a
pressureless fluid source, $w\left(  P\right)  =0$, while for the rest six
scaling solutions $w\left(  P\right)  =w\left(  P,\lambda,\kappa
,\sigma\right)  $, which can describe accelerated eras for for specific values
of the free parameters $\left\{  \lambda,\kappa,\sigma\right\}  $. Moreover,
the ninth exact cosmological solution which was found from the analysis of the
stationary points describes a de Sitter universe.

As far as the stability of the exact solutions at the stationary points is
concerned, seven of the points are always unstable. while only the set of the
points $B_{2}^{\pm}~$can be stable. Point $C_{2}$ which describes the de
Sitter universe, has one eigenvalue negative while the rest of the eigenvalues
are always negative. Consequently, according to the center manifold theorem we
found the internal surface where the point $C_{2}$ is a source. Moreover, in
the presence of additional matter source only additional tracking solutions
follow, similarly to the quintessence model.

From the above results we observe that the specific Chiral cosmological model
can describe the major eras of the cosmological history, that is, the late
expansion era, an unstable matter dominated era, and two scaling solutions
describe the radiation dominated era and the early acceleration epoch,
therefore, the model in terms of dynamics it is cosmologically viable.

From this analysis it is clear that the Chiral cosmological model can be used
as dark energy candidate. In a future work we plan to apply the cosmological
observations to constrain the theory.


\begin{thebibliography}{99}                                                                                               %


\bibitem {dataacc1}A. G. Riess, et al., Astron J. 116, 1009 (1998)

\bibitem {dataacc2}S. Perlmutter, et al., Astrophys. J. 517, 565 (1998)

\bibitem {data1}P. Astier et al., Astrophys. J. 659, 98 (2007)

\bibitem {data2}N. Suzuki et al., Astrophys. J. 746, 85 (2012)

\bibitem {Hinshaw:2012aka}G.~Hinshaw et al. [WMAP Collaboration],
Astrophys.\ J.\ Suppl.\ 208, 19 (2013)

\bibitem {Ade:2015xua}P.~A.~R.~Ade et al. [Planck Collaboration],
Astron.\ Astrophys.\ 594, A13 (2016)

\bibitem {guth}A. Guth, Phys. Rev. D 23, 347 (1981)

\bibitem {Ratra}B. Ratra and P.J.E. Peebles, Phys. Rev. D 37, 3406 (1988)

\bibitem {hor1}G.W. Hordenski, Int. J. Theor. Phys. 10, 363 (1975)

\bibitem {hor2}C. Deffayet, G.\ Esposito-Farese and A. Vikman, Phys. Rev. D
79, 084003 (2009)

\bibitem {hor3}A.A. Coley and R.J. van den Hoogen, Phys. Rev. D 62, 023517 (2000)

\bibitem {mod1}T. Clifton, P.G. Ferreira, A. Padilla and C. Skordis, Phys.
Rept. 513, 1 (2012)

\bibitem {mod2}J. Kluso\v{n}, Class. Quantum Grav. 28, 125025 (2011)

\bibitem {mod3}D. S\'{a}ez-G\'{o}mez, Phys. Rev. D 85, 023009 (2012)

\bibitem {sing}J.D. Barrow and A.A.H. Graham, Phys. Rev. D 91, 083513 (2015)

\bibitem {page}D.N. Page, Class. Quant. Grav. 1, 417 (1984)

\bibitem {udm}S. Basilakos and G. Luke-Gerakopoulos, Phys. Rev. D 78, 083509 (2008)

\bibitem {jdbnew}J.D. Barrow, Phys. Rev. D \textbf{48}, 1585 (1993)

\bibitem {muslinov}A. Muslimov, Class. Quant. Grav. \textbf{7,} 231 (1990)

\bibitem {ellis}G.F.R.\ Ellis and M.S. Madsen, Class. Quant. Grav. \textbf{8,}
667 (1991)

\bibitem {barrow1}J.D. Barrow and P. Saich, Class. Quant. Grav. \textbf{10,}
279 (1993)

\bibitem {newref2}R. de Ritis, G. Marmo, G. Platania, C. Rubano, P. Scudellaro
and C. Stornaiolo, Phys. Rev. D. \textbf{42} 1091 (1990)

\bibitem {ref001}S. Basilakos, M. Tsamparlis and A. Paliathanasis, Phys. Rev.
D 83, 103512 (2011)

\bibitem {ref002}J.D.\ Barrow and A. Paliathanasis, Phys.\ Rev. D 94, 083518 (2016)

\bibitem {cop}E.J. Copeland, \ M. Sami and S. Tsujikawa, IJMPD 15, 1753 (2006)

\bibitem {gen01}G. Leon and F.O. Franz Silva, Generalized scalar field
cosmologies, arXiv:1912.09856

\bibitem {ph1}V. Faraoni, Cosmology in Scalar-Tensor Gravity, Springer,
Dordrecht (2004)

\bibitem {ph2}V. Sivanesan, Phys. Rev. D \textbf{90}, 104006 (2014)

\bibitem {ph3}V. Gorini, A. Yu. Kamenshchik, U. Moschella and V. Pasquier,
Phys.\ Rev. D 69, 123512 (2004)

\bibitem {ph4}N.\ Chow and J. Khoury, Phys.\ Rev. D 80, 024037 (2009)

\bibitem {ph5}G. Leon and E.N. Saridakis, JCAP 1303, 025 (2013)

\bibitem {ph7}L.P. Chimento, M. Forte, R. Lazkoz and M.G. Richarte, Phys. Rev.
D 043502 (2009)

\bibitem {ph8}J. Socorro and E.O. Nunez, Eur. Phys. J. Plus 132, 168 (2017)

\bibitem {ph9}A. Giacomini, G. Leon, A. Paliathanais and S. Pan, EPJC 80, 184 (2020)

\bibitem {ph10}A. Paliathanasis, Gen. Rel. Grav. 51, 101 (2019)

\bibitem {ph11}D. Benisty and E.I. Guendelman, Class. Quantum Grav. 36, 095001 (2019)

\bibitem {hy1}A.D. Lindle, Phys. Rev. D \textbf{49}, 784 (1994)

\bibitem {hy2}E.J. Copeland, A.R. Liddle, D.H. Lyth, E.W. Steward and D.
Wands, Phys. Rev. D \textbf{49}, 6410 (1994)

\bibitem {hy3}S.A. Kim and A.R. Liddle, Phys. Rev. D \textbf{74}, 023513 (2006)

\bibitem {hy4}D. Wands, Lect. Notes Phys. \textbf{738}, 275 (2008)

\bibitem {atr1}P. Carrilho, D. Mulryne, J. Ronaye and T. Tenkanen, JCAP 06,
032 (2018)

\bibitem {atr3}P. Christodoulidis, D. Roest and E.I. Sfakianakis, JCAP 11, 002 (2019)

\bibitem {qq1}W. Hu, Phys. Rev. D \textbf{71}, 047301 (2005)

\bibitem {qq2}Y.-F. Cai, E.N. Saridakis, M.R. Setare and J.-Q. Xia,
Phys.\ Rept. \textbf{493}, 1 (2010)

\bibitem {qq3}R. Lazkoz, G. Leon and I. Quiros, Phys. Lett. B 649, 103 (2007)

\bibitem {atr6}S.V. Chervon, Quantum Matter \textbf{2}, 71 (2013)

\bibitem {atr7}I.V. Fomin, J. Phys.: Conf. Ser. 918, 012009 (2017)

\bibitem {sigm0}S. V. Ketov, Quantum Non-linear Sigma Models, Springer-Verlag,
Berlin, (2000).

\bibitem {andimakis}N. Dimakis, A. Paliathanasis, P.A.\ Terzis and T.
Christodoulakis, EPJC 79, 618 (2019)

\bibitem {per1}P. Christodoulidis, D. Roest and E.I. Sfakianakis, JCAP 12, 059 (2019)

\bibitem {2sfand}A. Paliathanasis and M. Tsamparlis, Phys.\ Rev. D 90, 043529 (2014)

\bibitem {Amendola:2006dg}L.~Amendola, G.~Camargo Campos and R.~Rosenfeld,
Phys.\ Rev.\ D 75, 083506 (2007)

\bibitem {Pavon:2007gt}D.~Pav\'{o}n and B.~Wang, Gen.\ Rel.\ Grav.\ 41, 1 (2009)

\bibitem {Chimento:2009hj}L.~P.~Chimento, Phys.\ Rev.\ D 81, 043525 (2010)

\bibitem {Arevalo:2011hh}F.~Arevalo, A.~P.~R.~Bacalhau and W.~Zimdahl,
Class.\ Quant.\ Grav.\ 29, 235001 (2012)

\bibitem {an001}A. Paliathanasis, S. Pan and W. Yang, IJMPD 28, 1950161 (2019)

\bibitem {an002}G. Papagiannopoulos, P. Tsiami, S. Basilakos and A.
Paliathanasis, EPJC 80, 55 (2020)

\bibitem {in1}D. Begue, C. Stahl and S.-S. Xue, Nucl. Phys. B 940, 312 (2019)

\bibitem {in2}M. Szydlowski, T. Stachowiak and R. Wojtak, Phys. Rev. D 73,
063516 (2006)

\bibitem {in3}W. Yang, S. Pan and A. Paliathanasis, MNRAS 482, 1007 (2019)

\bibitem {in4}S. Pan, W. Yang and A. Paliathanasis, to appear in MNRAS
(DOI:10.1093/mnras/staa213) (2020)

\bibitem {dyn1}L. Amendola, D.\ Polarski and S. Tsujikawa, IJMPD 16, 1555 (2007)

\bibitem {dyn2}G. Leon and E.N. Saridakis, JCAP 1504, 031 (2015)

\bibitem {dyn3}G. Leon, IJMPE 20, 19 (2011)

\bibitem {dyn4}T. Gonzales, G. Leon and I. Quiros, Class. Quantum Grav. 23,
3165 (2006)

\bibitem {dyn5}A. Giacomini,\ S. Jamal, G. Leon, A. Paliathanasis and
J.\ Saveedra, Phys.\ Rev. D 95, 124060 (2017)

\bibitem {dyn6}G. Chee and Y. Guo, Class. Quantum Grav. 29, 235022 (2012)
[Corrigendum: Class. Quantum Grav. 33, 209501 (2016)]

\bibitem {dyn7}S. Mishra and S.\ Chakraborty, EPJC 79, 328 (2019)

\bibitem {dyn8}H. Farajollahi and A. Salehi, JCAP 07, 036 (2011)

\bibitem {dyn9}M. Kerachian, G. Acquaviva and G. Lukes-Gerakopoulos,
Phys.\ Rev. D 101, 043535 (202)

\bibitem {cowei}S. Weinberg, Gravitation and cosmology: Principles and
applications of the general theory of relativity, Wiley, New York, (1972)

\bibitem {cob1}J.-Q. Xia, V. Vitagliano, S. Liberati, M. Viel, Phys. Rev. D.
85, 043520 (2012)
\end{thebibliography}
\end{document}